\documentclass[twoside,twocolumn,a4paper,11pt]{article}

\usepackage[a4paper,top=0.50cm,bottom=1.30cm,left=1.352cm,right=1.352cm]{geometry}

\usepackage{changepage}

\usepackage{xurl}
\usepackage[hidelinks,
            pdfauthor={Daniel Vallstrom},
            pdftitle={Cooperative Evolutionary Pressure and Diminishing Returns
                      Might Explain the Fermi Paradox: On What
                      Super-AIs Are Like}]{hyperref}

\usepackage{graphicx}

\usepackage[numbers, square, comma, compress]{natbib}
\usepackage{amssymb}
\usepackage{mathtools}
\usepackage{amsfonts}
\usepackage{amsthm}

\newcommand{\e}{\mathrm{e}}

\newtheorem{theorem}{Theorem}[section]

\theoremstyle{definition}
\newtheorem{defn}[theorem]{Definition}

\usepackage{mathdots}

\usepackage{algorithm}
\usepackage[noend]{algpseudocode}

\usepackage[misc]{ifsym}

\makeatletter
\newcommand{\algmargin}{\the\ALG@thistlm}
\makeatother
\algnewcommand{\parState}[1]{\State%
    \parbox[t]{\dimexpr\linewidth-\algmargin}%
              {\strut\hangindent=\algorithmicindent \hangafter=1 #1\strut}}

\usepackage{titlesec}
\titleformat{\section}      {\normalfont\large\bfseries}     {\thesection}      {0.8em}{}
\titleformat{\subsection}   {\normalfont\normalsize\bfseries}{\thesubsection}   {0.7em}{}
\titleformat{\subsubsection}{\normalfont\normalsize\itshape} {\thesubsubsection}{0.7em}{}
\titlespacing*{\section}      {0pt}{2.96ex plus 1ex minus .2ex}{1.59ex plus .2ex}
\titlespacing*{\subsection}   {0pt}{2.40ex plus 1ex minus .2ex}{0.90ex plus .2ex}
\titlespacing*{\subsubsection}{0pt}{2.14ex plus 1ex minus .2ex}{0.84ex plus .2ex}

\titleformat{\paragraph}    {\normalfont\normalsize\itshape} {\theparagraph}    {0.7em}{}
\titlespacing*{\paragraph}    {0pt}{1.85ex plus 1ex minus .2ex}{0.60ex plus .2ex}

\usepackage[font=small]{caption}

\usepackage{footnotebackref}

\usepackage[inline]{enumitem}

\usepackage{xcolor}

\usepackage{pgfplots}
\pgfplotsset{width=\linewidth,compat=1.18}

\usepgfplotslibrary{fillbetween}

\usetikzlibrary{patterns}
    
\usepackage{tikz}
\usetikzlibrary{arrows.meta,bending,positioning,arrows,decorations.markings}

\usepackage[color=white, hspace=16pt, icon=Note, voffset=-3pt, hoffset=3pt,
            author={Abstract}]{pdfcomment}

\usepackage{placeins}

\title{\vspace{1.18em}Cooperative Evolutionary Pressure and Diminishing Returns\linebreak[0]
                       Might Explain the Fermi Paradox:\linebreak[0]
                       On What Super-AIs Are Like}

\author{\vspace{0.9em}\hspace{0.25em}Daniel 
Vallstrom\textsuperscript{\href{mailto:daniel.vallstrom@gmail.com}
{\tiny\Letter}}\addtocounter{footnote}{1}\thanks{daniel.vallstrom@gmail.com}}

\date{\vspace{-0.0em}November 2026\vspace{-0.0em}}

\begin{document}

\if 01
\section{Title}
\fi
\maketitle

\if 01
\section{Abstract}
\fi
\abstract{
With an evolutionary approach,
the basis of
morality can be explained as 
adaptations to 
problems of cooperation.
With `evolution' taken in a broad sense,
AIs 
that satisfy
the conditions for evolution 
to apply
will be subject to the same
cooperative
evolutionary pressure as biological entities.
%
%
Here
the adaptiveness of increased cooperation as material safety and wealth increase
is discussed ---
for humans, for other societies, and for AIs.
%
%
Diminishing beneficial returns from 
increased access to material resources
also suggests the possibility that, 
on the whole,
there 
will be
no 
incentive
to for instance colonize entire galaxies,
thus providing
a possible 
explanation of the Fermi paradox,
wondering where everybody is.
%
It is further argued that
old societies 
could
engender
and eventually
give way to
super-AIs,
since it is likely that super-AIs are feasible, and fitter.
%
Closing
is
an aside on 
effective ways for 
morals and goals to affect life and society,
emphasizing 
evolutionary mismatches,
environments, 
cultures, and laws,
and 
exemplified 
by
how to eat.

%


%

`Diminishing returns' is defined, as less than roots,
the inverse of infeasibility.
It is also noted that there can be no exponential colonization or reproduction, 
for mathematical reasons, as each entity takes up a certain amount of space.
Appended are
an algorithm for colonizing for example a galaxy quickly,
and software for simulating signaling development.}
%






\section{Morality}

\subsection{Introduction: Morality as Evolutionary Solutions to Problems of Cooperation}
According to the evolutionary approach to morality (see
e.g.\ \cite{fit22, andre11, mann19, boehm12, krebs11, cosm18, toma18, baum13,
and22, baum16}),
\nocite{pri14}
morality 
evolves
as
solutions to
problems of cooperation:\footnote{There could possibly also be other types of morality.
For example, it is possible that feelings of
disgust to avoid pathogens,
or related phenomena,
could be adaptive, and count as morals
not directly tied to cooperation. 
Cf.\ \cite{fito22,fito23,oaten09disgust}.

We are interested in all actions
that are relevant, specifically to the Fermi 
paradox.
We 
don't have to restrict ourselves to
morality.
Still, we 
contend
that it is sufficient to 
focus on 
morality 
as solutions to problems of cooperation.
\nopagebreak[4]

Note too that morality as solutions to problems of cooperation does not comprise of one system,
but of many, at least for life on Earth.}
\nocite{rozin08disgust}
There is no good or bad \emph{a priori}. 
Instead, let $E$ be some entity: a biological being, a society, or an AI. 
Let $r$ and $c$ be changes to $E$, and let
$E_r$ and $E_c$ be the results, where $r$ made $E$ more 
recalcitrant
and 
callous,
while $c$ made $E$ more cooperative and considerate. 
For example, in the case where $E$ is biological, $r$ and $c$ can be mutations.
Then 
recalcitrance
and 
callousness
are
more pervasive
in $E_r$, while cooperation and considerateness are
less prevailing;
cooperation and considerateness are
more prevalent
in $E_c$, while 
recalcitrance
and 
callousness
are less
popular.

\subsubsection{Evolution Facilitates Cooperation}
\label{subsubsec:morex}
Moreover,
a main point
is that cooperativeness
is
often
adaptive.\cite{west2007evolutionary, sachs2004evolution, west2021ten,
lehmann2006evolution,
kurzban2015evolution,
clutton2009cooperation,
curry16, curry19b, curry19, fit22, andre11, boehm12,
cosm18, baum13, and22, 
baum16}\cite[\hspace{-0.15em}ch.\,11;\,pp.\,175--177,182--183]%
{krebs11}\cite{hoff16, mcn08, hall16,
mason16, bek09,
samu23,
gin01, mil02, bho19, farr18, sza21, rob21, oka20,
higgs2015rna, sudakow2024evolution}\footnote{Also 
cf.\ 
\nocite{alfano2024moral}
\hypertarget{sigSimRef}{\textit{evolutionary psychology}}
(see e.g.\ \cite{buss2019evolutionary, barkow1995adapted}),
\textit{culture}
(e.g.\ \cite{var17, mes11, 
baumard2025ecological, 
baumard2023gene, 
henrich2016secret, richerson2008not,
baumard2026evolved}),
\textit{signaling}
(\cite{sza23, sza22, bark19, sza21, rob21, gin01}$^{\ref{fn:sigSim}}$),
\nocite{mca18}
\textit{reputation}
(\cite{andre11, fit22, mil02, sza21, rob21, and22, rockenbach2006efficient}),
\textit{indirect reciprocity}
(\cite{rob21, oka20, nowak1998evolution, nowak2005evolution}),
\textit{punishment}
(\cite{fehr2000cooperation, fehr2004third, jordan2016third, fehr2004social,
balliet2011reward, gachter2008long, rand2009positive, fehr2002altruistic,
henrich2006costly, kurzban2007audience, barclay2006reputational,
brandt2003punishment, de2004neural, ostrom1992covenants, kamei2020group,
gurerk2006competitive,
jin2024institutions,
gelfand2024norm,
rockenbach2006efficient,
fitouchi2023punitive}),
\nocite{boyd2003evolution}
\textit{reciprocal altruism}
(\cite{curry08, trivers1971evolution}),
\textit{social norms}
(\cite{tankard2016norm, sparkman2017dynamic, gelfand2024norm, fehr2004social}),
\textit{assortment} (networks, evolutionary graph theory)
(\cite{ohtsuki2006simple}\cite[secs.\ on networks]{nowak2006five}%
\cite{eshel1982assortment,
      apicella2019evolution, samu23, dyble2016networks,
      lieberman2005evolutionary}),
\textit{institutions}
(\cite{lie2024social, powers2016institutions,
ostrom1990governing, ostrom2009understanding, 
north1990institutions, acemoglu2005institutions,
gelfand2024norm}),
\nocite{acemoglu2001colonial, acemoglu2012nations, acemoglu2002reversal}
%
%
and e.g.\ \cite{kap07, stan18, 
fito22, fito23, henrich21,
fitouchi2025prosocial, axelrod1981evolution,
kaznatcheev2019computational, 
benabou2006incentives,
brewer2007importance,
fehr2003nature,
kandori1992social,
civilini2024explosive, chen2025cooperation}.
See also \textit{evolutionary game theory} (e.g.\ \cite{SGame3}), 
which helps explain why 
cooperation
come about,
for instance
between cells,\cite{west2007evolutionary}
and between rats \cite{mason16}.
Cf.\ as well \textit{evolutionary game theory and ethics} (see e.g.\ \cite{SGame}).}
\nocite{de2014empirical}
\nocite{kimura1983neutral}
\nocite{fehr1999theory}
%
%
%
To continue the example, suppose that $h_c$ and $h_c'$ are two hunter-gatherers from $E_c$
above, while $h_r$ and $h_r'$ stem from $E_r$. Assume that they are all roughly equally capable 
foragers.
Say that they all go
foraging, each covering a separate part of the available land.
Suppose that their haul differ greatly, by pure chance: $h_c$ and $h_r$ get big hauls, with 
diminishing benefits, while $h_c'$ and $h_r'$ find nothing. $h_c$ shares her
haul with $h_c'$ while $h_r$ doesn't share with $h_r'$, 
with the result that
$h_c'$ survives while $h_r'$ dies.
Then after a while $h_c$ and $h_r$ get unlucky, finding nothing, while $h_c'$ gets lucky.
$h_c'$ shares her haul with $h_c$, 
with the result that
both $h_c$ and $h_c'$ survive again,
while $h_r$ dies.
(Cf.\,\cite{jaeggi2013natural, dyble2016networks, winterhalder1986diet,
kaplan2009evolutionary, powers2016institutions}%
\cite[\hspace{-0.18em}e.g.\hspace{0.18em}chs.\hspace{0.13em}7,\hspace{0.13em}10]%
{boehm12}\cite[\hspace{-0.15em}p.\,68]{baum13}%
\cite[\hspace{-0.15em}\S\,4.1]{and22};
for example, 
Hadza large-game hunters succeeded on average once every $37$ hunting days 
in the wet season \cite{hawkes1991hunting}.)

\subsection{A Definition of Diminishing Returns as Less than Roots}
\label{subsec:dimRetDef}


%
%
%
%

A definition of `diminishing', in the sense of eventually 
`inconsequential'
or `insignificant', 
follows from the definition of `feasible':

\subsubsection{Feasibility}

Something `feasible'
takes 
no more than
polynomial time, $x^n$.
`Infeasible' then takes more time. Typically, in practice, that means
at least exponential time, e.g.\ $\e^x$. 
However,
e.g.\ $x^{\ln x}$ also grows faster than polynomials,
but slower than exponentials.
(See the ordo notation too;
$f\!\in\!O(g)$, here, means 
\(\exists c\exists x_0 \, \forall x\!\!\geq\!\!x_0 \, f(x)\leq c\cdot g(x)\).)

The definition of feasible as taking no more than polynomial time 
\cite{papadimitriou2003computational, papadimitriou1994computational, 
cook2007overview, cobham1965intrinsic}\cite[sec.\,2]{edmonds1965paths}
works very well,
and is 
generally considered
correct,\cite{papadimitriou2003computational, papadimitriou1994computational, 
cook2007overview}\cite[pp.\,6--9]{garey1979computers}
in the sense that e.g.\ the definition of
computable as computable by a Turing machine is correct,\cite{soare1996computability}
or in the sense that the $\epsilon$-$\delta$ definition of continuity is correct.


\subsubsection{A Definition of Diminishing Returns}

Correspondingly then, something with 
(intrinsically) 
`diminishing' returns grows 
slower than the inverses of the polynomials, i.e.\ slower than 
root functions, $x^{1/n}$, $\sqrt[n]{x}$.
Typically, in practice, that means logarithmically, 
$\ln x$, or slower.
But theoretically, e.g.\ $(x^{\ln x})^{\langle-1\rangle}$,
the inverse of $x^{\ln x}$, also 
has diminishing returns, 
and grows faster than logarithms.
In other words, $f$ has diminishing returns, if and only if
something taking $f^{\langle-1\rangle}$ time is infeasible,
where $f^{\langle-1\rangle}$ exists. 
(See figure~\ref{fig:orders}.)

\begin{figure}
  \centering

  \begin{tikzpicture}
    \begin{axis}[
      legend style={font=\small},
      width = 1.0\linewidth,
      height = 40/39*1.0\linewidth,
      axis y line = left,
      axis x line = bottom,
      axis line style = {very thick},
      grid style = {thin, densely dotted, black!50},
      extra x ticks={1,5,15,25,35},
      xmin=1, xmax=40,
      ymin=0, ymax=40,
      xlabel = $x$,
      ylabel = {\(f(x)\)}]

      \addplot[smooth, domain=1:4, samples=100]{exp(x)};
      \addlegendentry{\(\e^x\)}
      
      \addplot[smooth, domain=1:7, samples=100, very thick]{x^ln(x)};
      \addlegendentry{\(x^{\ln x}\)}
      
      \addplot[smooth, domain=1:12, samples=100, densely dashed]{x^(3/2)};
      \addlegendentry{\(x^{3/2}\)}

      \addplot[smooth, domain=1:40, samples=100]{x^(100/99)};
      \addlegendentry{\(x^{100/99}\)}

      \addplot[
        domain=1:40,
        smooth,
        very thick,
        samples=100]
        {x};
      \addlegendentry{\(x\)}

      \addplot[smooth, domain=1:40, samples=100]{x - 1/exp(x)};
      \addlegendentry{\(x - 1/\e^x\)}
      
      \addplot[smooth, domain=1:40, samples=100]{x^0.99};
      \addlegendentry{\(x^{99/100}\)}

      \addplot[smooth, domain=1:40, samples=100, densely dashed]{x^(2/3)};
      \addlegendentry{\(x^{2/3}\)}

      \addplot[smooth, domain=1:7, samples=100, very thick]({x^ln(x)},x);
      \addlegendentry{\((x^{\ln x})^{\langle-1\rangle}\)}

      \addplot[smooth, domain=1.01:1.99, samples=10]{ln(x)/ln(2)};
      \addlegendentry{\hspace{-3.29em}{\large$\cdot$}\hspace{3.02em}\(\mathfrak{d}\,x\)}

      \addplot[smooth, domain=1:40, samples=100, densely dashdotted]{(ln(x))+(sin(deg(x)))};
      \addlegendentry{\(\ln x + \sin x\)}

      \addplot[smooth, domain=1:40, samples=100]{ln(x)};
      \addlegendentry{\(\ln x\)}

      \addplot[smooth, domain=1:40, samples=100, densely dotted, thick]{((-0.1)*(x*x)) + 4*x};
      \addlegendentry{\(-0.1 x^2 + 4x\)}

      \addplot[only marks, mark size=0.7pt, color=black]
              coordinates {(2,0.807354922058)};
      \addplot[smooth, domain=2.01:3.99, samples=20]{ln(x)/ln(2)};      

      \addplot[only marks, mark size=0.7pt, color=black]
              coordinates {(4,2.0397799235)};
      \addplot[smooth, domain=4.01:7.99, samples=30]{ln(x)/ln(2)};      
      \addplot[only marks, mark size=0.7pt, color=black]
              coordinates {(8,2.99538913702)};
      \addplot[smooth, domain=8.01:15.99, samples=40]{ln(x)/ln(2)};      
      \addplot[only marks, mark size=0.7pt, color=black]
              coordinates {(16,4.00001388293)};
      \addplot[smooth, domain=16.01:31.99, samples=50]{ln(x)/ln(2)};      
      \addplot[only marks, mark size=0.7pt, color=black]
              coordinates {(32,5.00000000022)};
      \addplot[smooth, domain=32.01:40, samples=40]{ln(x)/ln(2)};

    \end{axis}
  \end{tikzpicture}

  \vspace{-0.55em}
  \caption[ln x, ln x + sin x, d x, the inverse of x\^{}ln x, and
           -0.1x\^{}2 + 4x have diminishing 
           returns.]{$\ln x$, $\ln x + \sin x$,
                     $\mathfrak{d}\,x$ (see sec.\,\ref{subsubsec:dimRetEcon}),
                     $(x^{\ln x})^{\langle-1\rangle}$, 
                     and \(-0.1 x^2 + 4x\)
                     have diminishing returns.
                     \(x - 1/\e^x\) has decreasing returns and is hidden behind $f(x) = x$.}
  \label{fig:orders}
  \vspace{-0.10em}
\end{figure}

\begin{defn}[$DR$]
\label{def:DR}
$f$ has diminishing returns 
if
$\forall n\, f\in O(x^{1/n})$.
\end{defn}
Let $DR$ be the set of functions with diminishing returns;
$f\not\in DR \,\Leftrightarrow\, \exists n\, f \not\in O(x^{1/n})$.

See appendix \ref{appx:DRAlt} for alternative definitions,
using an integral-$O$ notation,
and further discussion.

\subsubsection{Definition Motive}

To 
motivate 
definition \ref{def:DR},
and why the diminishing returns concept is the inverse of infeasibility,
if something, $f$, 
has diminishing returns,
then no amount of extra feasible work, $g$, will push the marginal returns up
to significance, since 
$g\circ f$
has diminishing returns still
(i.e., duplicating the result of $f$ a feasible amount of $g$ times,
results in an insignificant marginal return 
still):
\begin{theorem}
\label{theorem:DRClosed}
If $f\in DR$, and $g\in O(x^k)$, 
then $g\circ f \in DR$.
\end{theorem}
\vspace{-1.0em}
\begin{proof}
Assume that $f\in DR$, and $g\in O(x^k)$.
Let $n\in \mathbb{N}$.
But $f\in O(x^{1/(kn)})$. Hence $g\circ f \in O(x^{1/n})$.
\end{proof}

For the other direction,
assume that $f$ doesn't have 
diminishing returns according to definition \ref{def:DR}.
Then there is an $n$ such that
$f \not\in O(x^{1/n})$.
Therefore
$x^n \circ f \not\in O(x)$.
Hence,
decreasing returns of $f$ can, in some sense, 
be overcome by a feasible amount of extra work,
by duplicating the result of $f$ $x^n$ times,
with the result $x^n\circ f$,
for
recurring
above unit 
marginal returns.
So $f$ does in fact have 
non-diminishing returns also in some intuitive sense.

It's good to have the concept of
(intrinsically) diminishing returns 
closed under 
natural operations, in particular feasible processes.
Regardless, definition \ref{def:DR} of diminishing returns 
as less than roots 
is the one used 
throughout this text,
rather than any economics definition.

\subsubsection{Diminishing Returns in Economics}
\label{subsubsec:dimRetEcon}

The concept of 
`diminishing returns' is used in economics. 
(Cf.\ e.g.\ 
\cite{brue1993retrospectives,
shephard1974law}\cite[sec.\ Diminishing Marginal Utility]{kenrick2009deep}.)
However, 
in particular,
the 
neoclassical 
treatment there doesn't suit
our purposes here.
For example, if `diminishing' is interpreted as just decreasing,
then `diminishing returns' could be arbitrary close to
constant
returns,
e.g.\ $x^{c}$ with $c\!<\!1$ close to $1$, 
or $h_1(x) = x - 1/\e^x$, or $h_2(x) = x - 1/\e^{\e^x}$, or 
$h_n(x) = x - 1/\e^{\e^{\iddots^x}}$ with $n$ iterated exponentiations,
which is not what we want
(e.g., 
$h_1'' = -\e^{-x}$, 
$h_2'' = (1 - \e^{x}) \e^{x - \e^{x}}$). 
(You also might want to call it `decreasing' if
you mean decreasing and not diminishing.)
You further don't want to require that returns will be non-increasing 
for them to be diminishing ---
that would be too strong and would exclude e.g.\ $\ln$.
(See figure~\ref{fig:orders}.)

%
%

There are also functions that intuitively have diminishing returns,
and have diminishing returns according the
definition used here, but where the returns are not decreasing 
in a mathematical sense, and 
hence do not have diminishing
returns according to any economics definition requiring decrease.
For example, let $\mathfrak{d}(x) = \log_2(x)$ except for $x = 2^n$
where $\mathfrak{d}(x) = \log_2(x-1/2^x) + 1/2^x$.
Then $\mathfrak{d}$ has diminishing returns, but the returns are not decreasing;
the same is true of $\ln x + \sin x$.
See figure~\ref{fig:orders}.

Introductory economics texts,
when exemplifying the concept of diminishing marginal utility,
seem to typically use quadratic 
functions, that eventually
have negative returns, 
e.g.\ \(-0.1 x^2 + 4x\) 
which has negative returns for $x>20$
(see figure~\ref{fig:orders}).\cite{todorova2021diminishing} 
And when the example functions don't 
have negative returns, they are typically logarithms, it 
seems.\cite{todorova2021diminishing}


Historically too, at first, 
`diminishing' meant diminishing, and not
merely decreasing, to some at least;\cite{brue1993retrospectives}
in neoclassical economics, `diminishing' is defined as
merely decreasing \cite{horowitz2007test}.

When testing marginal utility, 
econ\-o\-mists found logarithmic returns,
not merely decreasing 
re\-turns.\cite{horowitz2007test}
%

While we, here, are after some intrinsic property,
economists might be more interested in the immediate.
Still,
to have 
e.g.\
$9^{9^9}\!x - 1/\e^{\e^x}$
have diminishing returns,
while
e.g.\
$\ln x + \sin x$
doesn't,
perhaps 
in part
for 
mathematical expedience,
might not be ideal.


\subsection{Diminishing Returns Facilitate Cooperation}
\label{subsec:dimRet}
Importantly,
there are often 
diminishing,
logarithmic, beneficial returns from material 
resources \cite{oneill18, 
fanning2022social}.
%
For example, 
O'Neill, 
Fanning 
et al.\
\cite{oneill18}
found
that wealthy nations are 
often
past the ``turning point'' 
of the logarithmic curve of the benefit of material resources,
after which
using
more resources 
adds
very little to human well-being.
%
%
On an individual level,
well-being and life satisfaction have been found to depend logarithmically on 
income.\cite{kill21,stev13,stev08,kill23,owidhappiness,pinker18}
This diminishing effect  
facilitates
cooperation
as 
instead of using some resource yourself, for 
little or no gain,
you can 
reciprocally
give or trade it away,
like 
the cooperative hunter-gatherers
$h_c$ and $h_c'$ in section \ref{subsubsec:morex}
(cf.\ \cite[\hspace{-0.15em}\S\,4.1]{and22}).
%


%
Similar diminishing, logarithmic, returns have been found for research, with
findings
that research productivity declines,\cite{blo20}
%
and that research becomes decreasingly
disruptive and 
increasingly
narrow \cite{park23}
(cf.\,\cite{cowen19}).

Hard, or intractable or infeasible, computational problems also show
diminishing returns from using more resources, of course, by definition.

Animals, 
including humans, rats, and pigeons, 
have evolved super-exponential, hyperbolic, delay 
dis\-count\-ing.\cite{odum2011delay}
This hyperbolic discounting 
could be
consistent with 
diminishing returns from material resources:
Take an amount of resources, $\e^{x+1}$. Say that $\e^{x+1}$ is valued
only slightly more than $\e^{x}$, $1$ unit more, say.
If the resource or reward $\e^{x+1}$ is deferred to
later,
there will be 
a further
depreciation in
the evaluation due to the delay and uncertainty,
making the whole evaluation hyperbolic, say. 
%
%

Possibly relevant for e.g.\ colonization,
trying to 
ensure an outcome by duplicating processes or entities 
pursuing
the outcome also shows diminishing returns:
if you want an outcome $A$, by way of 
independent entities or processes $a_i$, with each 
$a_i$
ensuring $A$
with probability $p$, 
then duplicating instances $a_i$
has logarithmic returns since
with $n$ instances $a_i$ the probability of $A$ is $1 - (1-p)^n$.


%
%
For further discussions,
see sections \ref{subsubsec:predProj} and \ref{subsubsec:dimRet}, 
appendix \ref{sec:evoMod}, 
and 
\cite[sec.\ Diminishing Marginal 
Utility]{kenrick2009deep}\cite{brue1993retrospectives, horowitz2007test}.
\subsubsection{Predictions, Projections, Evidence}
\label{subsubsec:predProj}
One
prediction is then,
given the 
predisposition to cooperation
(cf.\ \cite{hamlin2011young, warneken2006altruistic}),
that cooperation will increase as
for example
material safety and wealth
increase,
especially
past
the ``turning point''
of
the logarithmic curve of the benefit of material resources,
\interfootnotelinepenalty=10000
ceteris paribus.\footnote{This 
assumes external variables remain constant. 
For instance, changes in the media environment that promote misinformation and divisiveness 
could counteract.
(Cf.\ 
\cite{lor22, bak21, ep23, koz20,
allcott2017social, vosoughi2018spread,
robertson2024morality}\cite[ch.\,5]{sing22}\cite{kup22,
charlesworth2022patternsB, 
starbird2019disinformation,
piccardi2025reranking}\cite[sec.\,Media]{gelfand2024norm}.)}
\interfootnotelinepenalty=1000
\nocite{bayer2020social}
%

Since the start of the industrial revolution 
there has been an exponential increase in GDP per capita
(from yearly increases), 
amounting to about a 
\mbox{1400\hskip 0.07em\%}
increase for the world
since then.\cite{OWiDGDPPC} See figure \ref{fig:gdp}.

\begin{figure}
  \centering

  \begin{tikzpicture}

    \begin{axis}[
      legend style={font=\small},
      width = 1.0\linewidth,
      height = 1.0\linewidth,
      axis y line = left,
      axis x line = bottom,
      axis line style = {very thick},
      xticklabel style={/pgf/number format/1000 sep={}},
      legend pos=north west,
      grid style = {thin, densely dotted, black!50},
      xmin=1820, xmax=2022,
      ymin=800, ymax=57000,
      xlabel = {year},
      ylabel = {GDP per capita}]

      \addplot+ [mark size=0.3] 
        table[x=Year,y=GDP per capita,col sep=comma] {gdp-per-capita-maddison-0-w-offshoots.csv}; 
      \addlegendentry{USA, CAN, AUS, NZL}

      \addplot+ [mark size=0.3]
        table[x=Year,y=GDP per capita,col sep=comma] {gdp-per-capita-maddison-1-w-euro.csv}; 
      \addlegendentry{Western Europe}

      \addplot+ [mark size=0.3] 
        table[x=Year,y=GDP per capita,col sep=comma] {gdp-per-capita-maddison-2-e-asia.csv}; 
      \addlegendentry{East Asia}

      \addplot+ [mark size=0.3] 
        table[x=Year,y=GDP per capita,col sep=comma] {gdp-per-capita-maddison-3-e-euro.csv}; 
      \addlegendentry{Eastern Europe}

      \addplot+ [mark size=0.3pt]
        table[x=Year,y=GDP per capita,col sep=comma] {gdp-per-capita-maddison-4-mid-e-n-africa.csv}; 
      \addlegendentry{Middle East, N Africa}
      
      \addplot+ [mark size=0.3] 
        table[x=Year,y=GDP per capita,col sep=comma] {gdp-per-capita-maddison-5-world.csv}; 
      \addlegendentry{World}
      
      \addplot+ [mark size=0.3]
        table[x=Year,y=GDP per capita,col sep=comma] {gdp-per-capita-maddison-6-latin-am.csv}; 
      \addlegendentry{Latin America}
      
      \addplot+ [mark size=0.3] 
        table[x=Year,y=GDP per capita,col sep=comma] {gdp-per-capita-maddison-7-s-se-asia.csv}; 
      \addlegendentry{S and SE Asia}
           
      \addplot+ [mark size=0.3] 
        table[x=Year,y=GDP per capita,col sep=comma] {gdp-per-capita-maddison-8-sub-sah-africa.csv}; 
      \addlegendentry{Sub Saharan Africa}

    \end{axis}
  \end{tikzpicture}

  \vspace{-0.55em}
  \caption[GDP per capita, from 1820 to 2022.]{GDP per capita, from 1820 to 2022,
           in 2011 international dollars. 
          Data from \cite{bolt2023maddison}, via \cite{OWiDGDPPC}.}
  \label{fig:gdp}
  \vspace{-0.10em}
\end{figure}


As a first check
of the prediction, 
we can look at already studied 
attitudes
that purportedly
or possibly
correlate with cooperativeness.
\interfootnotelinepenalty=10000
For example, 
the last few hundred years,
humans 
seemingly
have been
getting 
decreasingly
sexist, racist, homophobic, speciesist, 
and violent,
and 
increasingly
environmentally conscious.\cite{WVS, welzel13, ing18, ing05, %
OWiD, 
pinker11, pinker18}\footnote{Cf.\
\cite{cir16} 
where the authors were not able to state statistically that war casualties per capita have
declined over time;
war violence differs from interpersonal 
violence.}\cite[\hspace{-0.15em}ch.\,7]{mar17}\cite{ceci23, schaerer2023trajectory,
ceci2014women, birk21, card23, mccarthy22, mccarthy2021us,
charlesworth2022patterns, charlesworth2019patterns, charlesworth2022patternsB,
eagly2020gender, leach2023word, garg2018word,
vDem,
owid-women-rights,
owidHomo,
owidWPEI, sundstrom2017women,
unGII, owidGII}
\interfootnotelinepenalty=1000
%
\nocite{pew2020us}
%
%
\hypertarget{raceFigRef}{See}
figures \ref{fig:wEmpow}, \ref{fig:homoA}, \ref{fig:homoB}, \ref{fig:homoC},
\ref{fig:homicide},
and \ref{fig:prog}.
See also supplementary figures \ref{fig:race} and \ref{fig:homicideRecent} at the end.
%
%
%
%
%
%
%
%
%
%
%
%
(Cf.\ \cite{mast23} 
providing evidence
that people,
all over the world,
mistakenly
believe
that there is moral decline,
in part at least due to
evolved
negativity bias (on negativity bias, see e.g.\ \cite[esp.\ sections 3 and 1]{vaish2008not});
cf.\ \cite{baumeister2001bad, soroka2015news}.)


\begin{figure}
  \centering

  \begin{tikzpicture}

    \begin{axis}[
      legend style={font=\small},
      width = 1.0\linewidth,
      height = 1.0\linewidth,
      axis y line = left,
      axis x line = bottom,
      axis line style = {very thick},
      xticklabel style={/pgf/number format/1000 sep={}},
      legend pos=north west,
      xmin=1900, xmax=2023,
      ymin=0.0, ymax=1.0,
      xlabel = {year},
      ylabel = {Women's political empowerment index}]

      \addplot+ [name path=lower, fill=none, mark=none, draw=none, forget plot] table [
        x=Year, y=*Europe CI (Low), col sep=comma] {Women_political_empowerment_index.csv};
      \addplot+ [name path=upper, fill=none, mark=none, draw=none, forget plot] table [
        x=Year, y=*Europe CI (High), col sep=comma] {Women_political_empowerment_index.csv};
      \addplot+ [fill opacity=0.05, forget plot] fill between[of=lower and upper];

      \addplot+ [mark size=0.2] 
        table[x=Year,y=*Europe,col sep=comma] {Women_political_empowerment_index.csv}; 
      \addlegendentry{Europe}

      \addplot+ [name path=lower, fill=none, mark=none, draw=none, forget plot] table [
        x=Year, y=*Americas CI (Low), col sep=comma] {Women_political_empowerment_index.csv};
      \addplot+ [name path=upper, fill=none, mark=none, draw=none, forget plot] table [
        x=Year, y=*Americas CI (High), col sep=comma] {Women_political_empowerment_index.csv};
      \addplot+ [fill opacity=0.05, forget plot] fill between[of=lower and upper];

      \addplot+ [mark size=0.2] 
        table[x=Year,y=*Americas,col sep=comma] {Women_political_empowerment_index.csv}; 
      \addlegendentry{Americas}

      \addplot+ [name path=lower, fill=none, mark=none, draw=none, forget plot] table [
        x=Year, y=*Oceania CI (Low), col sep=comma] {Women_political_empowerment_index.csv};
      \addplot+ [name path=upper, fill=none, mark=none, draw=none, forget plot] table [
        x=Year, y=*Oceania CI (High), col sep=comma] {Women_political_empowerment_index.csv};
      \addplot+ [fill opacity=0.05, forget plot] fill between[of=lower and upper];

      \addplot+ [mark size=0.2] 
        table[x=Year,y=*Oceania,col sep=comma] {Women_political_empowerment_index.csv}; 
      \addlegendentry{Oceania}

      \addplot+ [name path=lower, fill=none, mark=none, draw=none, forget plot] table [
        x=Year, y=*World CI (Low), col sep=comma] {Women_political_empowerment_index.csv};
      \addplot+ [name path=upper, fill=none, mark=none, draw=none, forget plot] table [
        x=Year, y=*World CI (High), col sep=comma] {Women_political_empowerment_index.csv};
      \addplot+ [fill opacity=0.05, forget plot] fill between[of=lower and upper];

      \addplot+ [mark size=0.2] 
        table[x=Year,y=*World,col sep=comma] {Women_political_empowerment_index.csv}; 
      \addlegendentry{World}

      \addplot+ [name path=lower, fill=none, mark=none, draw=none, forget plot] table [
        x=Year, y=*Asia CI (Low), col sep=comma] {Women_political_empowerment_index.csv};
      \addplot+ [name path=upper, fill=none, mark=none, draw=none, forget plot] table [
        x=Year, y=*Asia CI (High), col sep=comma] {Women_political_empowerment_index.csv};
      \addplot+ [fill opacity=0.05, forget plot] fill between[of=lower and upper];

      \addplot+ [mark size=0.2] 
        table[x=Year,y=*Asia,col sep=comma] {Women_political_empowerment_index.csv}; 
      \addlegendentry{Asia}

      \addplot+ [name path=lower, fill=none, mark=none, draw=none, forget plot] table [
        x=Year, y=*Africa CI (Low), col sep=comma] {Women_political_empowerment_index.csv};
      \addplot+ [name path=upper, fill=none, mark=none, draw=none, forget plot] table [
        x=Year, y=*Africa CI (High), col sep=comma] {Women_political_empowerment_index.csv};
      \addplot+ [fill opacity=0.05, forget plot] fill between[of=lower and upper];

      \addplot+ [mark size=0.2] 
        table[x=Year,y=*Africa,col sep=comma] {Women_political_empowerment_index.csv}; 
      \addlegendentry{Africa}

      \addplot+ [name path=lower, fill=none, mark=none, draw=none, forget plot] table [
        x=Year, y=*Middle East and North Africa (MENA) CI (Low), col sep=comma]
        {Women_political_empowerment_index.csv};
      \addplot+ [name path=upper, fill=none, mark=none, draw=none, forget plot] table [
        x=Year, y=*Middle East and North Africa (MENA) CI (High), col sep=comma]
        {Women_political_empowerment_index.csv};
      \addplot+ [fill opacity=0.05, forget plot] fill between[of=lower and upper];

      \addplot+ [mark size=0.2] 
        table[x=Year,y=*Middle East and North Africa (MENA),col sep=comma]
             {Women_political_empowerment_index.csv}; 
      \addlegendentry{MENA}

    \end{axis}
  \end{tikzpicture}

  \vspace{-0.55em}
  \caption[Women's political empowerment index, from 1900 to 2023]{Women's
           political empowerment index, from 1900 to 2023.
           The index includes
           civil liberties,
           participation, and representation.\cite{sundstrom2017women} 
           Data from \cite[v2x\_gender]{vDem}. See also \cite{owidWPEI}.
           `MENA': Middle East and North Africa.}
  \label{fig:wEmpow}
  \vspace{-0.10em}
\end{figure}

%
%
%
%
%
%
%

\newcounter{homoFigLabelBaseValue}
\stepcounter{figure}
\setcounter{homoFigLabelBaseValue}{\value{figure}}          
\renewcommand{\thefigure}{{\thehomoFigLabelBaseValue}a}        

\begin{figure}
  \centering

  \begin{tikzpicture}

    \begin{axis}[
      legend style={font=\footnotesize},
      legend style={fill=none},  
      legend cell align=left,
      width = 0.55\linewidth,
      height = 0.79\linewidth,
      axis y line = left,
      axis x line = bottom,
      axis line style = {very thick},
      xticklabel style={/pgf/number format/1000 sep={}},
      yticklabel={$\pgfmathprintnumber{\tick}\%$},
      legend style={at={(axis cs:2023,-2.60)},anchor=south west},
      grid style = {thin, densely dotted, black!50},
      xmin=1993, xmax=2022,
      ymin=0, ymax=100,
      xlabel = {year},
      ylabel = {don't want homosexual neighbors}]

      \addplot+ [mark size=0.3] 
        table[x=Year,y=Neighbors being homosexuals: Mentioned,col sep=comma]
             {share-of-people-saying-they-do-not-want-homosexual-neighbors-1-12-8.csv}; 
      \addlegendentry{\texttt{CHN -01.1\,pp -02\%}}

      \addplot+ [mark size=0.3] 
        table[x=Year,y=Neighbors being homosexuals: Mentioned,col sep=comma]
             {share-of-people-saying-they-do-not-want-homosexual-neighbors-1-12-3.csv}; 
      \addlegendentry{\texttt{BLR -12.4\,pp -16\%}}

      \addplot+ [mark size=0.3] 
        table[x=Year,y=Neighbors being homosexuals: Mentioned,col sep=comma]
             {share-of-people-saying-they-do-not-want-homosexual-neighbors-1-12-5.csv}; 
      \addlegendentry{\texttt{BGR -07.9\,pp -12\%}}

      \addplot+ [mark size=0.3] 
        table[x=Year,y=Neighbors being homosexuals: Mentioned,col sep=comma]
             {share-of-people-saying-they-do-not-want-homosexual-neighbors-b4.csv}; 
      \addlegendentry{\texttt{HUN -38.4\,pp	-51\%}}

      \addplot+ [mark size=0.3] 
        table[x=Year,y=Neighbors being homosexuals: Mentioned,col sep=comma]
             {share-of-people-saying-they-do-not-want-homosexual-neighbors-1-12-11.csv}; 
      \addlegendentry{\texttt{EST -37.3\,pp -51\%}}

      \addplot+ [mark size=0.3] 
        coordinates {(1993,51.73824)(2022,29.193838)};
      \addlegendentry{\texttt{WLD -22.5\,pp -44\%}}

      \addplot+ [mark size=0.3] 
        table[x=Year,y=Neighbors being homosexuals: Mentioned,col sep=comma]
             {share-of-people-saying-they-do-not-want-homosexual-neighbors-1-12-7.csv}; 
      \addlegendentry{\texttt{CHL -29.7\,pp -52\%}}

      \addplot+ [mark size=0.3] 
        table[x=Year,y=Neighbors being homosexuals: Mentioned,col sep=comma]
             {share-of-people-saying-they-do-not-want-homosexual-neighbors-1-12-9.csv}; 
      \addlegendentry{\texttt{CZE -32.4\,pp -63\%}}

      \addplot+ [mark size=0.3] 
        table[x=Year,y=Neighbors being homosexuals: Mentioned,col sep=comma]
             {share-of-people-saying-they-do-not-want-homosexual-neighbors-1-12-12.csv}; 
      \addlegendentry{\texttt{FIN -12.9\,pp -51\%}}

      \addplot+ [mark size=0.3] 
        table[x=Year,y=Neighbors being homosexuals: Mentioned,col sep=comma]
             {share-of-people-saying-they-do-not-want-homosexual-neighbors-1-12-2.csv}; 
      \addlegendentry{\texttt{AUT -32.7\,pp -75\%}}

      \addplot+ [mark size=0.3] 
        table[x=Year,y=Neighbors being homosexuals: Mentioned,col sep=comma]
             {share-of-people-saying-they-do-not-want-homosexual-neighbors-1-12-6.csv}; 
      \addlegendentry{\texttt{CAN -19.5\,pp -66\%}}

      \addplot+ [mark size=0.3] 
        table[x=Year,y=Neighbors being homosexuals: Mentioned,col sep=comma]
             {share-of-people-saying-they-do-not-want-homosexual-neighbors-1-12-1.csv}; 
      \addlegendentry{\texttt{ARG -30.3\,pp -78\%}}

      \addplot+ [mark size=0.3] 
        table[x=Year,y=Neighbors being homosexuals: Mentioned,col sep=comma]
             {share-of-people-saying-they-do-not-want-homosexual-neighbors-1-12-10.csv}; 
      \addlegendentry{\texttt{DNK -09.5\,pp -81\%}}

    \end{axis}
  \end{tikzpicture}

  \vspace{-0.55em}
  \caption[Share of people saying that they don't want homosexual neighbors, 1993 to 2022.]
          {Share of people saying that they don't want homosexual neighbors, 1993 to 2022,
           for all 36 countries with entries for both 1993 and 2022.
           Wld means average of these 36 countries. 
           Legend also shows percentage point difference between 2022 and 1993,
           and relative difference. 
           Data from IVS (Integrated Values Surveys, i.e.\
           European Values Study (EVS) and World Values Survey (WVS) \cite{WVS}), via \cite{owidHomo}. 
           See \cite{owidHomo}.
           See figures \ref{fig:homoB} and \ref{fig:homoC} for rest of countries.
          }
          
  \label{fig:homoA}

  \vspace{0.20em}
\end{figure}

\if 01

ARG -30.3 pp -78%
AUT -32.7 pp -75%
BGR -07.9 pp -12%
BLR -12.4 pp -16%
BRA -23.4 pp -78%
CAN -19.5 pp -66%
CHL -29.7 pp -52%
CHN -01.1 pp -02%
CZE -32.4 pp -63%
DEU -26.8 pp -78%
DNK -09.5 pp -81%
ESP -17.4 pp -58%
EST -37.3 pp -51%
FIN -12.9 pp -51%
FRA -16.4 pp -68%
GBR -26.8 pp -86%
HUN -38.4 pp -51%
ISL -18.1 pp -90%
ITA -27.4 pp -70%
JPN -42.2 pp -62%
KOR -16.2 pp -17%
LTU -29.2 pp -33%
LVA -40.8 pp -52%
MEX -37.5 pp -62%
NGA +12.6 pp +17%
NLD -09.0 pp -75%
NOR -15.9 pp -82%
POL -42.6 pp -60%
PRT -37.5 pp -76%
ROU -21.7 pp -29%
RUS -14.4 pp -18%
SVK -25.9 pp -40%
SVN -13.3 pp -31%
SWE -15.3 pp -87%
TUR -16.0 pp -17%
USA -25.9 pp -67%

1993:
( 38.92215 + 43.33873 + 67.50484 + 79.01478 + 30.21665 + 29.70438 + 57.46667 + 71.9     + 51.30235 + 34.31254 + 11.65049 + 30.16553 + 72.91667 + 25.17007 + 24.3513  + 31.12231 + 75.27528 + 20.08547 + 39.22493 + 68.54599 + 95.84333 + 87.4     + 78.40531 + 60.22208 + 76.42358 + 12.02495 + 19.45117 + 70.46843 + 49.62414 + 75.43064 + 80.52014 + 64.04494 + 42.51208 + 17.66953 + 91.74757 + 38.59761 )  /  36
= 51,73824

2022:
( 8.572754 + 10.62448 + 59.57557 + 66.63367 + 6.776723 + 10.22335 + 27.72196 + 70.78629 + 18.90136 + 7.475639 + 2.161226 + 12.74079 + 35.62286 + 12.30045 + 7.906352 + 4.278382 + 36.84836 + 1.945659 + 11.83632 + 26.38581 + 79.59839 + 58.16014 + 37.59653 + 22.70626 + 89.04611 + 3.008684 + 3.512447 + 27.90773 + 12.15302 + 53.7121  + 66.1309  + 38.11105 + 29.18441 + 2.380118 + 75.7764  + 12.67587 )  /  36
= 29,193838
\fi

\renewcommand{\thefigure}{{\thehomoFigLabelBaseValue}b}        

\begin{figure}
  \centering

  \begin{tikzpicture}

    \begin{axis}[
      legend style={font=\footnotesize},
      legend style={fill=none},  
      legend cell align=left,
      width = 0.55\linewidth,
      height = 0.79\linewidth,
      axis y line = left,
      axis x line = bottom,
      axis line style = {very thick},
      xticklabel style={/pgf/number format/1000 sep={}},
      yticklabel={$\pgfmathprintnumber{\tick}\%$},
      legend style={at={(axis cs:2023,-2.60)},anchor=south west},
      grid style = {thin, densely dotted, black!50},
      xmin=1993, xmax=2022,
      ymin=0, ymax=100,
      xlabel = {year},
      ylabel = {don't want homosexual neighbors}]

      \addplot+ [mark size=0.3] 
        table[x=Year,y=Neighbors being homosexuals: Mentioned,col sep=comma]
             {share-of-people-saying-they-do-not-want-homosexual-neighbors-b1.csv}; 
      \addlegendentry{\texttt{NGA +12.6\,pp	+17\%}}

      \addplot+ [mark size=0.3] 
        table[x=Year,y=Neighbors being homosexuals: Mentioned,col sep=comma]
             {share-of-people-saying-they-do-not-want-homosexual-neighbors-b2.csv}; 
      \addlegendentry{\texttt{LTU -29.2\,pp	-33\%}}

      \addplot+ [mark size=0.3] 
        table[x=Year,y=Neighbors being homosexuals: Mentioned,col sep=comma]
             {share-of-people-saying-they-do-not-want-homosexual-neighbors-b3.csv}; 
      \addlegendentry{\texttt{LVA -40.8\,pp	-52\%}}

      \addplot+ [mark size=0.3] 
        coordinates {(1993,51.73824)(2022,29.193838)};
      \addlegendentry{\texttt{WLD -22.5\,pp -44\%}}

      \addplot+ [mark size=0.3] 
        table[x=Year,y=Neighbors being homosexuals: Mentioned,col sep=comma]
             {share-of-people-saying-they-do-not-want-homosexual-neighbors-b5.csv}; 
      \addlegendentry{\texttt{JPN -42.2\,pp	-62\%}}

      \addplot+ [mark size=0.3] 
        table[x=Year,y=Neighbors being homosexuals: Mentioned,col sep=comma]
             {share-of-people-saying-they-do-not-want-homosexual-neighbors-b6.csv}; 
      \addlegendentry{\texttt{MEX -37.5\,pp	-62\%}}

      \addplot+ [mark size=0.3] 
        table[x=Year,y=Neighbors being homosexuals: Mentioned,col sep=comma]
             {share-of-people-saying-they-do-not-want-homosexual-neighbors-c8.csv}; 
      \addlegendentry{\texttt{ESP -17.4\,pp	-58\%}}

      \addplot+ [mark size=0.3] 
        table[x=Year,y=Neighbors being homosexuals: Mentioned,col sep=comma]
             {share-of-people-saying-they-do-not-want-homosexual-neighbors-b7.csv}; 
      \addlegendentry{\texttt{ITA -27.4\,pp	-70\%}}

      \addplot+ [mark size=0.3] 
        table[x=Year,y=Neighbors being homosexuals: Mentioned,col sep=comma]
             {share-of-people-saying-they-do-not-want-homosexual-neighbors-b9.csv}; 
      \addlegendentry{\texttt{DEU -26.8\,pp	-78\%}}

      \addplot+ [mark size=0.3] 
        table[x=Year,y=Neighbors being homosexuals: Mentioned,col sep=comma]
             {share-of-people-saying-they-do-not-want-homosexual-neighbors-1-12-4.csv}; 
      \addlegendentry{\texttt{BRA -23.4\,pp -78\%}}

      \addplot+ [mark size=0.3] 
        table[x=Year,y=Neighbors being homosexuals: Mentioned,col sep=comma]
             {share-of-people-saying-they-do-not-want-homosexual-neighbors-b10.csv}; 
      \addlegendentry{\texttt{NOR -15.9\,pp	-82\%}}

      \addplot+ [mark size=0.3] 
        table[x=Year,y=Neighbors being homosexuals: Mentioned,col sep=comma]
             {share-of-people-saying-they-do-not-want-homosexual-neighbors-b11.csv}; 
      \addlegendentry{\texttt{NLD -09.0\,pp	-75\%}}

      \addplot+ [mark size=0.3] 
        table[x=Year,y=Neighbors being homosexuals: Mentioned,col sep=comma]
             {share-of-people-saying-they-do-not-want-homosexual-neighbors-b12.csv}; 
      \addlegendentry{\texttt{ISL -18.1\,pp	-90\%}}

    \end{axis}
  \end{tikzpicture}

  \vspace{-0.55em}
  \caption[Share of people saying that they don't want homosexual neighbors, 1993 to 2022.]
          {Share of people saying that they don't want homosexual neighbors, 1993 to 2022.
           See figure \ref{fig:homoA} caption.
           See figures \ref{fig:homoA} and \ref{fig:homoC} for rest of countries.
          }
  \label{fig:homoB}
  \vspace{0.20em}
\end{figure}

\renewcommand{\thefigure}{{\thehomoFigLabelBaseValue}c}        

\begin{figure}
  \centering

  \begin{tikzpicture}

    \begin{axis}[
      legend style={font=\footnotesize},
      legend style={fill=none},  
      legend cell align=left,
      width = 0.55\linewidth,
      height = 0.79\linewidth,
      axis y line = left,
      axis x line = bottom,
      axis line style = {very thick},
      xticklabel style={/pgf/number format/1000 sep={}},
      yticklabel={$\pgfmathprintnumber{\tick}\%$},
      legend style={at={(axis cs:2023,-2.60)},anchor=south west},
      grid style = {thin, densely dotted, black!50},
      xmin=1993, xmax=2022,
      ymin=0, ymax=100,
      xlabel = {year},
      ylabel = {don't want homosexual neighbors}]

      \addplot+ [mark size=0.3] 
        table[x=Year,y=Neighbors being homosexuals: Mentioned,col sep=comma]
             {share-of-people-saying-they-do-not-want-homosexual-neighbors-c1.csv}; 
      \addlegendentry{\texttt{KOR -16.2\,pp	-17\%}}

      \addplot+ [mark size=0.3] 
        table[x=Year,y=Neighbors being homosexuals: Mentioned,col sep=comma]
             {share-of-people-saying-they-do-not-want-homosexual-neighbors-c2.csv}; 
      \addlegendentry{\texttt{TUR -16.0\,pp	-17\%}}

      \addplot+ [mark size=0.3] 
        table[x=Year,y=Neighbors being homosexuals: Mentioned,col sep=comma]
             {share-of-people-saying-they-do-not-want-homosexual-neighbors-c3.csv}; 
      \addlegendentry{\texttt{RUS -14.4\,pp	-18\%}}

      \addplot+ [mark size=0.3] 
        table[x=Year,y=Neighbors being homosexuals: Mentioned,col sep=comma]
             {share-of-people-saying-they-do-not-want-homosexual-neighbors-c4.csv}; 
      \addlegendentry{\texttt{ROU -21.7\,pp	-29\%}}

      \addplot+ [mark size=0.3] 
        table[x=Year,y=Neighbors being homosexuals: Mentioned,col sep=comma]
             {share-of-people-saying-they-do-not-want-homosexual-neighbors-c5.csv}; 
      \addlegendentry{\texttt{SVK -25.9\,pp	-40\%}}

      \addplot+ [mark size=0.3] 
        coordinates {(1993,51.73824)(2022,29.193838)};
      \addlegendentry{\texttt{WLD -22.5\,pp -44\%}}

      \addplot+ [mark size=0.3] 
        table[x=Year,y=Neighbors being homosexuals: Mentioned,col sep=comma]
             {share-of-people-saying-they-do-not-want-homosexual-neighbors-c6.csv}; 
      \addlegendentry{\texttt{SVN -13.3\,pp	-31\%}}

      \addplot+ [mark size=0.3] 
        table[x=Year,y=Neighbors being homosexuals: Mentioned,col sep=comma]
             {share-of-people-saying-they-do-not-want-homosexual-neighbors-c7.csv}; 
      \addlegendentry{\texttt{POL -42.6\,pp	-60\%}}

      \addplot+ [mark size=0.3] 
        table[x=Year,y=Neighbors being homosexuals: Mentioned,col sep=comma]
             {share-of-people-saying-they-do-not-want-homosexual-neighbors-c9.csv}; 
      \addlegendentry{\texttt{USA -25.9\,pp	-67\%}}

      \addplot+ [mark size=0.3] 
        table[x=Year,y=Neighbors being homosexuals: Mentioned,col sep=comma]
             {share-of-people-saying-they-do-not-want-homosexual-neighbors-c10.csv}; 
      \addlegendentry{\texttt{PRT -37.5\,pp	-76\%}}

      \addplot+ [mark size=0.3] 
        table[x=Year,y=Neighbors being homosexuals: Mentioned,col sep=comma]
             {share-of-people-saying-they-do-not-want-homosexual-neighbors-b8.csv}; 
      \addlegendentry{\texttt{FRA -16.4\,pp	-68\%}}

      \addplot+ [mark size=0.3] 
        table[x=Year,y=Neighbors being homosexuals: Mentioned,col sep=comma]
             {share-of-people-saying-they-do-not-want-homosexual-neighbors-c11.csv}; 
      \addlegendentry{\texttt{GBR -26.8\,pp	-86\%}}

      \addplot+ [mark size=0.3] 
        table[x=Year,y=Neighbors being homosexuals: Mentioned,col sep=comma]
             {share-of-people-saying-they-do-not-want-homosexual-neighbors-c12.csv}; 
      \addlegendentry{\texttt{SWE -15.3\,pp	-87\%}}

    \end{axis}
  \end{tikzpicture}

  \vspace{-0.55em}
  \caption[Share of people saying that they don't want homosexual neighbors, 1993 to 2022.]
          {Share of people saying that they don't want homosexual neighbors, 1993 to 2022.
           See figure \ref{fig:homoA} caption.
           See figures \ref{fig:homoA} and \ref{fig:homoB} for rest of countries.
          }
  \label{fig:homoC}
  \vspace{-0.50em}
\end{figure}

\setcounter{figure}{\value{homoFigLabelBaseValue}}
\renewcommand{\thefigure}{\arabic{figure}}

\hypertarget{trustFigRef}{Interpersonal} trust may correlate with cooperativeness, 
too.\cite{darcy2026trust, knack1997does, algan2014trust}
Figures \ref{fig:trust} and \ref{fig:trustB} show recent changes.

There is also 
evidence 
that cooperation increases with material
wealth
from e.g.\ 
ultimatum game experiments 
(where a random player $a$ gets
some pie $p$ (typically in the form of money)
to split between $a$ and $b$;
if $b$ rejects the split they both get nothing
\cite{guth13,cap21}).
For instance, 
unfair offers (cf.\ \cite{andre11})
are rejected less often as $p$ increases.\cite{and11,oost04}
(As wealth increases, 
the relative value of a pie $p$ decreases,
and cooperation increases.)
(Punishing low offers by rejecting them promotes cooperation.\cite{ball13} Cf.\ \cite{schr14}.)
In addition,
as $p$ increases,
\cite{and11,oost04} found that
offers 
as a share of $p$
decreased, but see 
\cite{larney2019stake}. 
In
dictator games (where $b$ cannot reject the split \cite{cap21}), 
the share of $p$ that $a$ offers decreases as $p$ increases
\cite{and11,engel2011dictator,larney2019stake};
cf.\,\cite{korn15,net11}.
People also, 
in 
certain
experiments,
get less reluctant to lie as incentives get 
bigger.\cite{gnee05, gibson2013preferences, conrads2014honesty,
kajackaite2017incentives, gerlach2019truth}
The cooperativeness of young American adults has increased 
for 61 years.\cite{yuan22}
And wealthy returned
misdelivered money more often than poor.\cite{andreoni17}
See also
\cite{korn15, net11, schr14, anderson2006non,
carpenter2007demand}\cite[\hspace{-0.2em}ch.\,5]{fito22}
--- e.g., the less costly a third party perceived a fixed cost punishment of
antisocial behavior to be, the more likely she was to carry out the prosocial 
punishment \cite{schr14} (cf.\ \cite{trax12}).
See 
life history theory
as well, 
where cooperation is linked to 
absence of early-life 
stress.\cite{lett20, gri11, akee18, wu20, kap15, net20, mccullough2013harsh,
kesternich2020early, ellis2009fundamental}
\nocite{deitzer2024harsh}
%
%
%
Cf.\ also links between wealth, patience, and 
cooperation.\cite{pepp17}\cite[\hspace{-0.2em}e.g.\,\S\,V.D]{beck97}\cite{epp20,
sun21, lie22, and22, curry08, boon2022optimal, boon2024effect}
See also e.g.\ \cite{henrich2005economic, gelfand2011differences,
falk2018global, sng2018behavioral, thielmann2020personality,
wu2025social, macchia2021link, darcy2026trust}.
%
\nocite{mani2013poverty}

\begin{figure}
  \centering

  \begin{tikzpicture}

    \begin{axis}[
      legend style={font=\footnotesize},
      legend style={fill=none},  
      legend cell align=left,
      width = 0.52\linewidth,
      height = 0.74\linewidth,
      axis y line = left,
      axis x line = bottom,
      axis line style = {very thick},
      xticklabel style={/pgf/number format/1000 sep={}},
      legend pos = outer north east,
      grid style = {thin, densely dotted, black!50},
      xmin=1525, xmax=2021,
      ymin=0, ymax=40,
      xlabel = {year},
      ylabel = {homicides per 100,000 people}]

      \addplot+ [name path=lower, fill=none, mark=none, draw=none, forget plot] table [
        x=year, y=lower, col sep=comma] {IHME-GBD_2021_DATA-e74d460c-1.csv};
      \addplot+ [name path=upper, fill=none, mark=none, draw=none, forget plot] table [
        x=year, y=upper, col sep=comma] {IHME-GBD_2021_DATA-e74d460c-1.csv};
      \addplot+ [fill opacity=0.05, forget plot] fill between[of=lower and upper];

      \addplot+ [mark size=0.3] 
        table[x=year,y=val,col sep=comma]
             {IHME-GBD_2021_DATA-e74d460c-1.csv}; 
      \addlegendentry{\texttt{Wrld -01.74 -26\%~~41y}}  

      \addplot+ [mark size=0.3] 
        table[x=Year,y=Rate,col sep=comma]
             {homicide-rates-across-western-europe-1.csv}; 
      \addlegendentry{\texttt{CoSa -26.10	-93\% 331y}}

      \addplot+ [mark size=0.3] 
        table[x=Year,y=Rate,col sep=comma]
             {homicide-rates-across-western-europe-2.csv};  
      \addlegendentry{\texttt{SwFi -14.20	-77\% 200y}}

      \addplot+ [mark size=0.3] 
        table[x=Year,y=Rate,col sep=comma]
             {homicide-rates-across-western-europe-3.csv};  
      \addlegendentry{\texttt{EnWa -03.70	-71\% 431y}}

      \addplot+ [mark size=0.3] 
        table[x=Year,y=Rate,col sep=comma]
             {homicide-rates-across-western-europe-4.csv};
      \addlegendentry{\texttt{BEL~~-12.48 -95\% 495y}}

      \addplot+ [mark size=0.3] 
        table[x=Year,y=Rate,col sep=comma]
             {homicide-rates-across-western-europe-5.csv};
      \addlegendentry{\texttt{FRA~~-19.55 -97\% 495y}}

      \addplot+ [mark size=0.3] 
        table[x=Year,y=Rate,col sep=comma]
             {homicide-rates-across-western-europe-6.csv};
      \addlegendentry{\texttt{NLD~~-35.25 -98\% 496y}}

      \addplot+ [mark size=0.3] 
        table[x=Year,y=Rate,col sep=comma]
             {homicide-rates-across-western-europe-7.csv};
      \addlegendentry{\texttt{ESP~~-12.00 -95\% 496y}}

      \addplot+ [mark size=0.3] 
        table[x=Year,y=Rate,col sep=comma]
             {homicide-rates-across-western-europe-8.csv};
      \addlegendentry{\texttt{DEU~~-08.55 -95\% 445y}}

      \addplot+ [mark size=0.3] 
        table[x=Year,y=Rate,col sep=comma]
             {homicide-rates-across-western-europe-9.csv};
      \addlegendentry{\texttt{CHE~~-06.29 -93\% 446y}}

      \addplot+ [mark size=0.3] 
        table[x=Year,y=Rate,col sep=comma]
             {homicide-rates-across-western-europe-10.csv};
      \addlegendentry{\texttt{ITA~~-38.32 -99\% 445y}}

      \addplot+ [mark size=0.3] 
        table[x=Year,y=Rate,col sep=comma]
             {homicide-rates-across-western-europe-11.csv};
      \addlegendentry{\texttt{IRL~~-05.74 -97\% 295y}}

    \end{axis}
  \end{tikzpicture}

  \vspace{-0.55em}
  \caption[Homicide rates, 1525 to 2021.]
          {Homicide rates, 1525 to 2021.
           Legend also shows change from first to last entry,
           relative difference, and time span. 
           Non-``Wrld'' data are from \cite{owidHomicideLongTerm};
           for original sources, see
           \cite{owidHomicideLongTerm}.
           Data for ``Wrld'' (world) are from \cite{ihme};
           see also \cite{owidHomicideRate}.
           See \cite{owidHomicideDataSources} too.
           `CoSa': Corsica and Sardinia,
           `SwFi': Sweden and Finland,
           `EnWa': England and Wales.
           For a zoom-in on recent years, see figure \ref{fig:homicideRecent}.
          }
  \label{fig:homicide}
  \vspace{-0.10em}
\end{figure}
%


You can also model how cooperation evolves under diminishing returns. 
Results from this modeling
similarly
indicate
that 
cooperation and fairness are facilitated by diminishing returns.
See appendix \ref{sec:evoMod}.



Artificial super-intelligences (super-AIs, ASIs)
would 
arguably
be under a similar 
cooperative evolutionary pressure as biological entities and societies.
\nopagebreak[3]
See ap\-pen\-dix~\ref{sec:appendixSuperAI}.

\subsection{The Speed of Evolution}
\label{subsec:speed}
Regarding the speed of evolution, more complex human biological evolutionary adaptations take 
thousands of years 
(hence 
for example evolutionary psychologists' emphasis on hunter-gatherer 
environments when studying human adaptations).\cite[\hspace{-0.15em}fn.\,3]{cosm18}

\begin{figure}
  \centering
  \includegraphics[width=\linewidth]{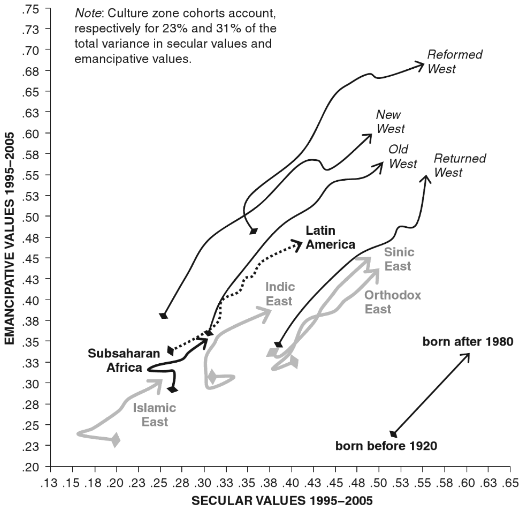}
  \vspace{-1.70em}
  \caption[Generational, Planckian value progression. Figure 2.5 from
            Christian Welzel \cite{welzel13}. (With permission.)]{Generational,
            Planckian 
            value progression. Figure 2.5 from Christian Welzel
            \cite{welzel13}.
            {\footnotesize (With permission.)}}
  \label{fig:prog}
  \vspace{-0.10em}
\end{figure}

The speed of human cultural evolution, 
in the sense of
information accumulation,
is,
ostensibly,
increasing rapidly.
For example, \cite{born14} found that scientific output is increasing exponentially, 
doubling every 9 years; see also \cite{larsen2010rate}.
On the other hand, research shows diminishing, logarithmic, returns.\cite{blo20,park23,cowen19}
There is also the question of the dissemination of information.
Besides, information alone is often not enough; environments, cultures, and laws are important
(see section \ref{subsubsec:env}).
Moreover, 
a significant
part of cultural evolution happens
generationally too,
because of 
relative
adult non-plasticity \cite{fawcett2015adaptive,ismail2017cerebral};
see e.g.\ \cite{welzel13,ing18}, and figure~\ref{fig:prog}
(but see also \cite{tor19}).


The
neural network based
AI AlphaZero achieved superhuman performance in for example chess and Go in 24 hours of learning 
from self-play, starting from scratch.\cite{alphaZero17}

%
%
%
%
%
%

\begin{figure}
  \centering

  \begin{tikzpicture}

    \begin{axis}[
      legend style={font=\footnotesize},
      legend style={fill=none},  
      legend cell align=left,
      width = 0.67\linewidth,
      height = 0.80\linewidth,
      axis y line = left,
      axis x line = bottom,
      axis line style = {very thick},
      xticklabel style={/pgf/number format/1000 sep={}},
      yticklabel={$\pgfmathprintnumber{\tick}\%$},
      legend style={at={(axis cs:2023,-1.60)},anchor=south west},
      grid style = {thin, densely dotted, black!50},
      xmin=1984, xmax=2022,
      ymin=0, ymax=80,
      xlabel = {year},
      ylabel = {share saying ``most can be trusted''}]

      \addplot+ [mark size=0.3] 
        coordinates {(1984,45.91667)(1993,55.53398)(2004,64.12512)(2010,74.91604)(2022,73.86647)};
      \addlegendentry{\texttt{DNK 27.9\,pp}}

      \addplot+ [mark size=0.3] 
        coordinates {(1984,55.47098)(1993,60.69411)(1998,64.77374)(2010,73.28255)(2022,72.09141)};
      \addlegendentry{\texttt{NOR 16.6\,pp}}

      \addplot+ [mark size=0.3] 
        coordinates {(1984,56.03191)(1993,59.52381)(1998,47.923)(2004,56.75382)(2010,59.9127)(2022,68.35999)};
      \addlegendentry{\texttt{FIN 12.3\,pp}}

      \addplot+ [mark size=0.3] 
        coordinates {(1984,38.08169)(1993,50.25825)(2004,59.38604)(2010,54.11947)(2014,66.1409)(2022,57.02047)};
      \addlegendentry{\texttt{NLD 18.9\,pp}}

      \addplot+ [mark size=0.3] 
        coordinates {(1984,46.28354)(1998,39.44921)(2010,45.57007)(2014,51.44999)(2022,48.53989)};
      \addlegendentry{\texttt{AUS \ 2.3\,pp}}

      \addplot+ [mark size=0.3] 
        coordinates {(1984,47.68692)(1993,50.65347)(2004,38.43174)(2010,41.79429)(2022,46.69302)};
      \addlegendentry{\texttt{CAN -1.0\,pp}}

      \addplot+ [mark size=0.3] 
        coordinates {(1984,37.9408905)(2010,40.105393)(2022,43.479468)};
      \addlegendentry{\texttt{WLD \ 5.5\,pp}}

      \addplot+ [mark size=0.3] 
        coordinates {(1984,25.92423)(1993,28.87377)(1998,32.13228)(2004,33.05191)(2010,35.58326)(2014,44.62337)(2022,41.61021)};
      \addlegendentry{\texttt{DEU \ 9.7\,pp}}

      \addplot+ [mark size=0.3] 
        coordinates {(1984,37.41689)(1993,37.58655)(1998,39.75332)(2004,39.64758)(2010,36.58759)(2014,35.93942)(2022,33.70288)};
      \addlegendentry{\texttt{JPN -3.7\,pp}}

      \addplot+ [mark size=0.3] 
        coordinates {(1984,35.97938)(1993,33.57314)(1998,30.26421)(2004,27.33333)(2010,28.00356)(2014,26.45815)(2022,32.93173)};
      \addlegendentry{\texttt{KOR -3.0\,pp}}

      \addplot+ [mark size=0.3] 
        coordinates {(1984,24.53229)(1993,32.75099)(2004,31.75)(2010,28.74554)(2022,26.55384)};
      \addlegendentry{\texttt{ITA \ 2.0\,pp}}

      \addplot+ [mark size=0.3] 
        coordinates {(1984,22.29873)(1993,21.35729)(2004,21.43032)(2010,23.68455)(2022,26.25627)};
      \addlegendentry{\texttt{FRA \ 4.0\,pp}}
      
      \addplot+ [mark size=0.3] 
        coordinates {(1984,24.53115)(1993,22.35529)(1998,17.08436)(2004,15.01535)(2010,17.36787)(2014,19.20537)(2022,19.18133)};
      \addlegendentry{\texttt{ARG -5.3\,pp}}

      \addplot+ [mark size=0.3] 
        coordinates {(1984,17.028)(1993,30.24167)(1998,26.42384)(2004,20.79574)(2010,15.44872)(2014,12.4)(2022,10.48835)};
      \addlegendentry{\texttt{MEX -6.5\,pp}}

    \end{axis}
  \end{tikzpicture}

  \vspace{-0.55em}
  \caption[Share saying that most people can be trusted, 1984 to 2022.]
          {Share saying that most people can be trusted, 1984 to 2022,
           for all 20 countries with entries for both 1984 and 2022.
           Wld means average of these 20 countries. 
           Legend also shows percentage point difference between 2022 and 1984.
           Data from IVS (EVS and WVS \cite{WVS}), via \cite{owidTrust}. 
           See \cite{owidTrust}.
           See figure \ref{fig:trustB} for rest of countries.
          }
          
  \label{fig:trust}

  \vspace{0.20em}
\end{figure}


\subsection{Enabling Morals and Goals to Affect Life}

A look at 
human moral progression 
the last few hundred years
suggests that 
humans 
have been getting
increasingly
cooperative and
\interfootnotelinepenalty=9900
considerate\footnote{`Considerateness' is used here as 
a trait
that 
in humans
ostensibly correlates with,
or signals,
cooperativeness 
(cf.\ \cite[e.g.\ table 5]{buss90}\cite{li2002necessities, buss1986preferences}),
but 
is not directly equivalent to 
cooperativeness 
itself.}
(see section \ref{subsubsec:predProj}).
\interfootnotelinepenalty=1000
%
%
(For a longer perspective, see \cite{kaplan2009evolutionary}
(cf.\,\cite{holden2003spread}).)
%
%

\subsubsection{Diminishing Returns}
\label{subsubsec:dimRet}
%
This 
considerateness
%
%
%
%
is made easier when 
the direct cost to ourselves is
small or zero
--- or, as in maybe the case of how to eat, 
when the direct cost is even negative
(appx.~\protect\ref{sec:mismatchNegCost}; sec.~\ref{subsec:dimRet}; appx.~\ref{sec:howToEat}).

%
%
%
%
%
%



\subsubsection{Creating Good Environments}
\label{subsubsec:env}

However,
%
we typically don't
deliberately
act according to what is considered good according to some more or less abstract ethics. 
Rather, our actions are largely shaped by our culture, environment, and habits,
which often have a more significant impact on our behavior than our deliberate
intentions.\cite{hof12, mil17, inz21, duck16, ent2015trait,
galla2015more,
hofmann2024going,
hofmann2024self,
sheeran2016intention,
bargh1999unbearable,
gollwitzer2006implementation,
wood2022habits}
\nocite{berg22}
\nocite{mazar2022illusory}
\nocite{cialdini2005basic}
\nocite{de2012taking}
\nocite{adriaanse2014effortless}
\nocite{de2024getting}
\nocite{hofmann2025self}
\nocite{gollwitzer1999implementation}
%
\nocite{webb2006does}
\nocite{rhodes2012experimental}
\nocite{sheeran2002intention}
\nocite{inzlicht2021integrating}
%
%
\nocite{ouellette1998habit}
\nocite{bargh1997automaticity}
%
\nocite{hofmann2009impulse}
\nocite{kahneman2011thinking}
%
%
\nocite{nordgren2009restraint}
%
%
%
For example, if you have access to 
palatable,
high-calorie,
non-satiating food, 
\hypertarget{veganDietExRef}{you}
tend to get overweight or obese,
regardless 
of 
your 
intentions;\cite{chow12, swin11, hall19, allcott2019food,
owidobesity}[appx.\,\ref{sec:howToEat}]
\nocite{Phelps2024WorldwideTI}
\nocite{mcgrosky2025energy}
if 
you have access to
only
low-calorie, 
nutrient-rich, satiating food, with a small environmental cost, you tend to eat healthy and 
have a small environmental impact,
and live 
longer \cite{will07};
\interfootnotelinepenalty=1900
the increased food supply is enough to explain all the weight gain,
and the increased waste \cite{chow12, swin11}.\footnote{Case in point:
The 
Okinawans had, up to WWii, access  
largely
to
vegetables,
including 
sweet potatoes,\cite{will07, willcox2009okinawan}
and,
according to \cite{will07, willcox2008they},
had e.g.\ 
exceptionally high
longevity and
number of centenarians per capita,
but cf.\ \cite{poulain2011exceptional}.
After WWii their food access gradually 
Americanized,\cite{todo04, nytoni04, will07, willcox2009okinawan}
but also Japanized,\cite{todo04, willcox2009okinawan}
and 2015
male Okinawans ranked 36th of 47 prefectures in life expectancy while the women ranked 
7th \cite{jap15} (cf.\,\cite{todo04, nytoni04}) ---
e.g., the calories
in some fatty and sugary mix (650 kcal/100 \nolinebreak[2]g say) are an order of 
magnitude more than in vegetables 
(say 25 kcal for spinach, 75 kcal for potatoes). Cf.\ appx.\,\ref{sec:howToEat}.}
\interfootnotelinepenalty=1000
%
%
%
%


Other areas work the same as the food case,
with greater 
supply or
access 
leading to,
or correlating with,
greater 
prevalence
(cf.\ \cite{hof12, mil17, inz21, duck16, ent2015trait}):
suicides,\cite{suic14, pinker18} 
gun violence,\cite{lopez2018america, gunsrev, hepburn2004firearm, studd22} 
car travel,\cite{cars}
opioid usage,\cite{ruhm18}
road usage \cite{litman,dur11}.

What we can do then is to see to it that the culture,
the environment,
laws,
and norms
are
such that they
steer us in the right 
direction.\cite{hof12, mil17, inz21, duck16, ent2015trait,
moodie2013profits, stuckler2012manufacturing, mialon2020overview, jahiel2007industrial,
stuckler2012big, chater2023frame}
As in the food 
case.\cite{moodie2013profits, stuckler2012manufacturing, stuckler2012big, chater2023frame,
will07, chow12, swin11, hall19}

In particular, it 
is beneficial 
to have
cultures, environments, and laws 
that 
facilitate cooperation,
for instance
through
feedback and
punishment
(cf.\ AI), 
including 
ways to
withhold
cooperation
with 
defectors
(cf.\ 
\cite{hauert2002volunteering, arai2023punish, rockenbach2006efficient}).
(See 
above; cf.\ \cite{herr08, mon07, ofosu2019same}.)
\nocite{paluck2012salience}
%
%
For example, with an evolutionary approach to morality, 
one could put less emphasis on liberty and no-censorship rights, 
and more emphasis on cooperation;
there would be no unassailable right 
to spread misinformation to anyone,
let alone
millions;
and there would be no unassailable right
to create media environments that
facilitate 
misinformation, division, or antisocial punishment.
(Cf.\ \cite{lor22, kozyreva2023resolving, pers20, bak21, ep23, koz20,
allcott2017social, vosoughi2018spread, robertson2024morality,
piccardi2025reranking,
penny21, penny19, kozyreva2024toolbox, 
blair2024interventions,
globig2023changing,
van2024social, baribi2024supersharers, allen2024quantifying, mccabe2024post,
jhaver2021evaluating, chandrasekharan2017you,
grinberg2019fake, 
robertson2024inside, farrell2024bias, kim2021distorting, lazer2018science,
brady2021social, 
jia2024embedding, 
moodie2013profits, stuckler2012manufacturing,
mialon2020overview, jahiel2007industrial, stuckler2012big, chater2023frame},
and e.g.\ 
``shouting `fire' in a crowded theater''.)
\nocite{brag22, rathje2021out, lorenz2020behavioural, brady2020mad,
crockett2017moral, suhay2018polarizing, robertson2024changing, 
gillespie2018custodians, allcott2020welfare, 
pennycook2021psychology, bor2022psychology, 
brady2023algorithm, yildirim2023short, 
frimer2023incivility, 
firth2019online,
torn22, 
haidtMedia, haidtMedia2, 
nguyen2025feeds,
keles2020systematic, huang2022meta,
haidt2026social}
%

See appendix \ref{sec:mismatchNegCost} for more.

\subsection{Feasibility Facilitates Cooperation}
\phantomsection
\label{subsec:feasibility}
\hypertarget{a_T}{There} 
is also the fact that any evolutionary moral solution has to be feasible, leading to
more general
moral solutions, as a practical matter.
For example,
let
\(a_\top\!\coloneqq \)
\textit{always do that which provides the highest fitness}.
Then $a_\top$ ought to be infeasible and
provide strictly higher fitness than anything actual.
Cf.\ \cite[sec.\ Consciousness: The Human Inclusive Fitness 
Maximizer?]{tooby1990past}\cite{cosmides1987evolution}.
All the while, old societies and super-AIs, and 
humans, 
would
perhaps
hold $a_\top$ immoral,
because $a_\top$ is evolutionary foreign,
unpredictable (cf.\hspace{0.19em}\cite{turp21,walk21}),
and potentially malefic;
with `old societies' here meaning societies 
older
than human society.%
%
%
%
\footnote{Still, 
in lieu of $a_\top$, 
events might have a distribution such that
the present predicts the future;
a biologically feasible adaptation
could be to in the future
be proficient at what one does in the present.}
%
%
%

\vspace{1.0em}
\section{The Fermi Paradox}
\vspace{0.1em}

\subsection{Introduction}
There is likely an abundance of planets where complex life could 
develop.\cite{gow11, line04, chop18, bryson2021occurrence}
For example, one simulation of galactic habitability, taking for instance supernovas into 
account, says that 0.3\% of all stars in 
the Milky Way host a habitable and tidally non-locked planet, assuming that the development of 
life, complex life, and ozone layers typically takes as much time as it did on Earth.\cite{gow11}
%
Planets supporting complex life for billions of years 
also ought to be possible;\cite{gow11, line04}
a further
simulation estimated that the median ``terrestrial like'' planet in the local volume,
i.e.\ within 35 million light-years, 
is around 7 billion years old around FGK stars and around 8 billion years 
old around M dwarfs.\cite{zack16}

Furthermore, that life, also complex life, develop doesn't seem that 
unlikely, perhaps.\cite{line02,bada04,chop18,miz22,bonner88,powell20}
(That life appears doesn't seem that unlikely, maybe;
see \cite{moody2024nature, mahendrarajah2023atp, betts18} too 
(cf.\ \cite{gob21, kras22, sha18, marz18, zhao2026extremophile, loeb_2014}).
\interfootnotelinepenalty=1000
See also
\cite{lah01, smith16, kipp20, mull22, whit22, xav22, xav20,
nowak2008prevolutionary, alakuijala2024computational, kruszewski2022emergence,
schmickl2016life,
gold1992deep,
flemming2019bacteria,
bar2018biomass}.\footnote{There are 
also
``exponentially'' small estimates of the likelihood that life starts (see e.g.\ \cite{hall12}).
Still, we
don't know how life started.\cite{walker_2017,powell20,wikiAbio}
Maybe
the fact that we don't know
ought not warrant
a minuscule
estimate of how likely it is that 
\pagebreak[2]
life starts.
See also e.g.\ \cite[sec.\ Cooperation in the RNA 
World]{higgs2015rna}\protect\nopagebreak[4]
for a discussion of some 
possible ways to avoid
exponentially small probabilities.
Cf.\ \cite{valiant2009evolvability} too,
which contains a discussion of feasible, i.e.\ polynomially bounded, evolution.
(Cf.\ maybe also that, from a theoretical perspective,
even if not directly applicable practically,
the existence of
self-reproducing entities or machines, outputting their own description, has a fairly short 
proof.\cite{yanofsky2003universal}
And \cite{alakuijala2024computational} found that self-replicators do emerge under fairly weak assumptions.)
Cf.\ as well \cite{mills2025reassessment}, reassessing proposed evolutionary 
``hard steps''.}
%
\nocite{wilf2010there}
\nocite{furukawa2025bio}
\nocite{gianni2026emergence}
%
And evolution can induce intelligence, under certain conditions, it would seem.
%
%
Note \nopagebreak[3] too 
that
cooperation,
maybe also facilitated by kin selection
(assortment),
might very well have facilitated 
early life evolution.\cite{higgs2015rna,west2021ten})
\interfootnotelinepenalty=1000

Of 521 astrobiologists answering a survey, 86.6\,\% agreed that
extraterrestrial life is likely, 
while 1.9\,\% disagreed.\cite{vickers2025surveys}


At first glance,
just a single society
should quickly, relative to the age of the universe, be able to 
engender
an exploration and a colonization of
for example its galaxy
with self-replicating ``probes''
(super-AIs)
(traveling at a fraction of the speed of light),
it would seem;
see appendix~\ref{sec:appendixColAlg}.\footnote{Although special relativity, for instance, places 
restrictions on fast travel, limiting heavier vessels to slower speeds, lighter vessels can 
theoretically travel fairly fast. According to
\cite{lubin16} ``it is within our technological reach'' to get spacecrafts weighing
a couple of
grams up to speeds around 0.1 of the speed of light, using laser arrays and light-sails.

    

(The diameter of the Milky Way is perhaps around 0.1 million light-years. The distance to the 
Andromeda galaxy is about 2.5 million light-years. 
There are maybe 0.1--0.4 trillion stars in the Milky Way, and 1 trillion in Andromeda.)
    
(Big Bang occurred 13.8 billion years ago and 
the Solar System formed 4.6 billion years ago.)}

The Fermi paradox is the apparent contradiction between the seemingly likely abundance of old 
societies
and super-AIs
throughout the universe and our lack of evidence of 
them.\cite{cirk18}
%
%

\subsection{On What Old Societies and Super-AIs Are Like}

\subsubsection{Cooperative Pressure and Diminishing Returns Could Explain the Fermi Paradox}
\label{subsubsec:FermiSolution}
The Fermi paradox is a paradox only if one assumes that old societies
or super-AIs
are
exhaustively
expansive.
And 
the above discussion suggests the possibility that 
that is precisely how they are not.

There are regularly diminishing
beneficial returns from material resources
(see section \ref{subsec:dimRet}).
If old societies 
and super-AIs
are
cooperative,
then they ought to need a
good reason in order to construct something as
impactful as a
\href{https://en.wikipedia.org/wiki/Kardashev_scale}{Kardashev type III}
society.
If, 
furthermore,
there are 
consistently
diminishing beneficial returns from material resources,
then you would expect there to be for example no Kardashev type III societies. 
And you would expect there to be no 
exhaustive
cosmological colonization.

Note 
also
that evolution does 
not
necessarily imply 
readiness
for ``exponential'' or even fast 
reproduction:

\enlargethispage{-1.0\baselineskip}

\paragraph{No exponential colonization or reproduction}
\label{para:NoExpCol}
First, 
entities living on a surface can at most spread or reproduce ``quadratically fast'',
as a function of time, 
for mathematical reasons,
as each entity takes up a certain amount of space and the travel for a new entity needs to be at 
least ``square root far''; only ``quadratically'' many children will ever be 
useful.
Cf.\ fn.\,\ref{fn:PNP}, appx.\,\ref{sec:appendixColAlg}
--- or consider the case with just one dimension instead of two or three; 
there it becomes 
clearer what's going on:
regardless of how many children each node spawns in the one dimensional case,
on average only exactly one child will ever be 
productive,
except for the root node in the bidirectional $\mathbb{Z}$ case. 
See figure \ref{fig:noExpRep}.

\begin{figure}
  \centering

  \begin{tikzpicture}
    
    

    \draw ( 0,0) node [minimum size=0.6cm,draw,double,circle,font=\footnotesize,inner sep=0pt] (a) {$0$};
    \draw ( 1,0) node [minimum size=0.6cm,draw,circle,font=\footnotesize,inner sep=0pt] (b) {$1$};
    \draw ( 2,0) node [minimum size=0.6cm,draw,circle,font=\footnotesize,inner sep=0pt] (d) {$2$};
    \draw ( 3,0) node [minimum size=0.6cm,draw,circle,font=\footnotesize,inner sep=0pt] (f) {$3$};
    
    \begin{scope}
      \clip (3.5,-0.32) rectangle (4.19,0.32);
      \draw ( 4,0) node [minimum size=0.6cm,draw,circle,font=\footnotesize,inner sep=0pt] (h) {$4$};
    \end{scope} 

    \draw (-1,0) node [minimum size=0.6cm,draw,circle,font=\footnotesize,inner sep=0pt] (c) {$-1$};
    \draw (-2,0) node [minimum size=0.6cm,draw,circle,font=\footnotesize,inner sep=0pt] (e) {$-2$};
    \draw (-3,0) node [minimum size=0.6cm,draw,circle,font=\footnotesize,inner sep=0pt] (g) {$-3$};

    \begin{scope}
      \clip (-3.5,-0.32) rectangle (-4.19,0.32);
      \draw (-4,0) node [minimum size=0.6cm,draw,circle,font=\footnotesize,inner sep=0pt] (i) {$-4$};
    \end{scope} 

    \draw[->] (a) to [bend left=30] (b);
    \draw[->] (b) to [bend left=30] (d);
    \draw[->] (d) to [bend left=30] (f);
    \draw[->] (f) to [bend left=30] (h);        

    \draw[->] (a) to [bend right=30] (c);
    \draw[->] (c) to [bend right=30] (e);
    \draw[->] (e) to [bend right=30] (g);
    \draw[->] (g) to [bend right=30] (i);

    \begin{scope}[decoration={
                  markings,
                  mark=at position 0.5 with {\arrow{>}}}
                 ] 

      \draw[->,-Rays,red,postaction={decorate}] (b) to [bend left=30] (a);
      \draw[->,-Rays,red,postaction={decorate}] (d) to [bend left=30] (b);
      \draw[->,-Rays,red,postaction={decorate}] (f) to [bend left=30] (d);    
      \draw[->,-Rays,red,postaction={decorate}] (h) to [bend left=30] (f);    
        
      \draw[->,-Rays,red,postaction={decorate}] (c) to [bend right=30] (a);
      \draw[->,-Rays,red,postaction={decorate}] (e) to [bend right=30] (c);
      \draw[->,-Rays,red,postaction={decorate}] (g) to [bend right=30] (e);    
      \draw[->,-Rays,red,postaction={decorate}] (i) to [bend right=30] (g);    
    \end{scope}


  \end{tikzpicture}

  \vspace{-0.30em}
  \caption[No exponential reproduction or colonization, 1D.]
          {No exponential reproduction or colonization, 1D.
           Red edges
           are unproductive.
           The root node is $0$.}
  \label{fig:noExpRep}
  \vspace{-0.10em}
\end{figure}

\enlargethispage{-1.0\baselineskip}

\paragraph{Fast reproduction can be maladaptive}
Second, 
evolution 
only
leads to fitter entities;
readiness
for very fast reproduction isn't guaranteed \nolinebreak[2]
--- 
high speeds could be
maladaptive;
\cite{koo14} found that 
animal 
species generally don't maximize reproduction.
%
And
environmental changes, 
cultural evolution,
affect phenotypes.
Cf.\ life history theory (e.g.\ \cite{ellis2009fundamental,kap15});
in particular, 
low morbidity and mortality,
higher availability of resources,
and high population density and social competition
predict slower strategies,
while
low morbidity and mortality,
higher availability of resources,
and low population density and social competition
predict faster strategies \cite{ellis2009fundamental,rotella2021increasing}.
%
%

This is consistent with 
the human 
fertility
rate, 
which continues to decrease; 
the number of births peaked 2012 and is projected to continue to get smaller, 
the number of children under 5 peaked 2017 and
is
projected to continue to get
smaller,
the number of people under 15 peaked 2020
and is projected to decrease,
and (hence) human population is projected to
decrease
within a 
\hypertarget{humanFertBFigRef}{few}
generations.\cite{owidpeakchild, owidfuturepopulationgrowth,
owidfertilityrate, owid-population-growth, vollset2020fertility, call23}
See figure \ref{fig:humanFertility}, and supplementary figures
\ref{fig:humanFertilityB} and \ref{fig:humanFertilityC} at the end.

\begin{figure}
  \centering

  \begin{tikzpicture}

    \begin{axis}[
      legend style={font=\footnotesize},
      legend style={fill=none},  
      legend cell align=left,
      width = 0.78\linewidth,
      height = 1.00\linewidth,
      axis y line = left,
      axis x line = bottom,
      axis line style = {very thick},
      xticklabel style={/pgf/number format/1000 sep={}},
      legend style={at={(axis cs:2024,0.60)},anchor=south west},
      grid style = {very thin, dashed, gray!40},
      grid = major,
      ytick distance=1,
      xmin=1950, xmax=2023,
      ymin=0, ymax=8,
      xlabel = {year},
      ylabel = {births per woman}]

      \addplot+ [mark size=0.6] 
        table[x=Year,y=Fertility_rate,col sep=comma]
             {children-born-per-woman-001-low.csv};
      \addlegendentry{Low-inc}

      \addplot+ [mark size=0.3] 
        table[x=Year,y=Fertility_rate,col sep=comma]
             {children-born-per-woman-000-afr.csv}; 
      \addlegendentry{Africa}

      \addplot+ [mark size=0.6] 
        table[x=Year,y=Fertility_rate,col sep=comma]
             {children-born-per-woman-110-ken.csv};
      \addlegendentry{Kenya}

      \addplot+ [mark size=0.3] 
        table[x=Year,y=Fertility_rate,col sep=comma]
             {children-born-per-woman-001-lowmid.csv};
      \addlegendentry{Lower-mid}

      \addplot+ [mark size=0.3] 
        table[x=Year,y=Fertility_rate,col sep=comma]
             {children-born-per-woman-111-wrld.csv};
      \addlegendentry{World}

%
%
%
%
%
%
%
%
%
%

      \addplot+ [mark size=0.3] 
        table[x=Year,y=Fertility_rate,col sep=comma]
             {children-born-per-woman-111-upmid.csv};
      \addlegendentry{Upper-mid}

      \addplot+ [mark size=0.3] 
        table[x=Year,y=Fertility_rate,col sep=comma]
             {children-born-per-woman-010-high.csv};
      \addlegendentry{High-inc}

%
%
%
%
%
%

      \addplot+ [green,mark size=0.3] 
        table[x=Year,y=Fertility_rate,col sep=comma]
             {children-born-per-woman-011-kor.csv};
      \addlegendentry{S Korea}

    \end{axis}
  \end{tikzpicture}

  \vspace{-0.55em}
  \caption[Human fertility rates, 1950 to 2023.]
          {Human fertility rates, 1950 to 2023.
           Data from \cite{UNFert}, via \cite{owidfertilityrate}.
           'Low-inc': low-income countries,
           'Lower-mid': lower-middle-income countries,
           'Upper-mid': upper-middle-income countries,
           'High-inc': high-income countries. See figures \ref{fig:humanFertilityB} and
           \ref{fig:humanFertilityC} for other countries and regions.}
  \label{fig:humanFertility}
  \vspace{-0.10em}
\end{figure}

\subsubsection{``Known Unknowns'' as Non-Solutions to the Fermi Paradox}
Regarding $\mathrm{CO_2}$ pollution as a possible, partial, explanation of the Fermi paradox,
societies according to the above evolve not only technologically or in terms of GDP
and the capacity for $\mathrm{CO_2}$ pollution should correlate with
general social
or cooperative
development, to some degree at least,
in particular according to the prediction in section \ref{subsubsec:predProj}.
Enough so that $\mathrm{CO_2}$ pollution ought to not play a significant part in the explanation
of the Fermi paradox;
the analogs and implicants
of dying off due to $\mathrm{CO_2}$ pollution
ought to not be
adaptive.
This seems to be corroborated by 
the human case, assuming that we are mediocre.
%
For similar reasons as in the $\mathrm{CO_2}$ pollution case, the risk of a society destroying 
nuclear war also ought to
not play a significant part in the explanation of the Fermi paradox.
Also similarly,
non-sustainability ought to not play a significant part in the explanation of the Fermi paradox 
either.
In sum,
the above problems are problems of cooperation, and cooperation is
central to evolution (see above, and specifically cf.\ \cite{curry19c}; 
cf.\ also e.g.\ \cite{hale2024long}; 
but cf.\ evolutionary mismatch (\ref{sec:solutions}) and e.g.\ \cite{tooby1990past} too).

Besides, the above type of problems ought to be of much lesser concern to super-AIs (see below).

Note that this 
section is not
about 
specific
likelihoods of these
``known known'' and
``known unknown''
risks, and doesn't try to
quantify or minimize them.
Rather,
the point is that,
given cooperative pressure,
any argument that there is a 
(mitigatable)
risk is an argument 
\emph{against} that that risk matters for the explanation of
the Fermi paradox.
Instead 
you would need ``unknown unknowns'' with 
high 
likelihoods, which ought to be
close to
self-contradictory --- and not only certain ``unknown unknowns'' for
humans 
and other biological entities
but for super-AIs 
as well.


\subsubsection{Cooperative Cliques}
With cooperative pressure, super-AIs and old societies could form information
sharing cliques
(in the graph theoretical sense).
Then there also would be 
no reason for any AI or society in a clique to duplicate explorations.
%
%

\subsubsection{On Old Societies or Super-AIs Reaching Out}
As for old societies or super-AIs contacting us, they ought to get more or less 
no benefit from it.
Moreover, whether to reach out and help someone falls on a continuous spectrum:
As a result of cooperative evolutionary pressure,
an old society or a super-AI probably would help 
if, 
say, the cost was small and
another old society or super-AI
asked for assistance. 
On the other hand, an old society or a super-AI probably won't try to intervene in everything
that is going on in the universe ---
that would
be a sort of utilitarian philosophical misconception,
at least according to the therapeutic approach to philosophy (see 
section \ref{subsubsec:utilitarianism},
or cf.\ \cite[last section (pp.\ 312--313)]{hoff16}
and
\cite[\S\S\ 3, \hspace{-0.08em}28]{cosm18}).
For example, we probably wouldn't fault old societies or super-AIs for not helping, say, the 
dinosaurs or the Neanderthals.

\subsubsection{From Biological to Artificial}
Old societies could
engender
and eventually
give way to
some efficient, 
reproducing 
artificial or machine beings 
--- super-AIs ---
as super-AIs ought to be safer, more efficient, more flexible, fitter.
The 
questions are if super-AIs are feasible, and if they are fitter.
And
the answer to both questions is likely ``yes''; see appendix~\ref{sec:appendixSuperAI}.

\subsection{Learning from Old Societies and Super-AIs}

%
Let's assume that
\begin{enumerate*}[label=(\roman*)]
\item
\label{firstPremise}
it is likely
that old societies or super-AIs exist or have existed, and
that
\item
we
give credence to the
proposition
that old societies and super-AIs have evolved also morally.
\end{enumerate*}
With these premises 
we can get moral suggestions by observing the non-actions of old societies
and super-AIs.

\subsubsection{No Type III Societies, No Exhaustive Colonization}
For example, 
given the premises,
old societies and super-AIs
don't seem to build societies of
\href{https://en.wikipedia.org/wiki/Kardashev_scale}{Kardashev type III},\cite{griffith2015g, garr15}
with its large environmental impact, so we should probably not aim to do that
either.\footnote{There have also been other searches 
for some other artifacts,
for example Dyson spheres in our 
neigh\-bor\-hood,
not finding any.\cite[p.\ \hspace{-0.12em}250]{cirk18}\cite{sua22}}
They
also don't seem to colonize 
exhaustively,
so we should also probably not aim to do that either.

\subsubsection{Utilitarianism Non-Espoused}
\label{subsubsec:utilitarianism}
According to the therapeutic approach to philosophy e.g., 
utilitarianism is a 
scientistic
armchair
misconception.%
%
\cite{horwich13NYT,horwich12Book}
%
There is also
domain specific and scientific criticism of utilitarianism,
see e.g.\ Hoffman et al.\ \cite[last section (pp.\ 312--313)]{hoff16}, 
Cosmides et al.\ \cite[\S\S\ 3, \hspace{-0.08em}28]{cosm18},
and \cite{turp21,walk21}.
See also Andr{\'e} et al.\ \cite[\S\S\ 5.2,\,6.1]{and22}.

Moreover, you can conclude that 
utilitarianism
is seemingly 
mistaken,
given the premises,
also by observing that
it appears to not be espoused by old societies or super-AIs. Because if it was,
then there ought to be a colonization of Earth, or 
interventions in events on Earth,
to alleviate suffering and increase happiness
here.

\subsubsection{No Bad AI}
Similarly, one can draw tentative conclusions about what will work in AI.
The symbolic approach to AI (see e.g.\ \cite{mitch19}) is based more on rules and discontinuous 
functions.
If
the
symbolic
approach
can produce AI that is at odds with what we observe, for example AI that could treat
the Milky Way
purely as a resource 
and do things that we would be able to observe,
%
%
then 
that
speaks against the fruitfulness of that
rule based and discontinuous 
approach,
given the first premise \ref{firstPremise}.
All the while,
the approach to AI that is based on neural networks
is
consistent with our observations (see appx.~\ref{sec:appendixSuperAI}).


\subsubsection{No High-End Fast Life History Strategy}
In a similar manner,
and very tentatively, the Fermi paradox could inform life history theory,
assuming super-AIs exist:
Super low morbidity and mortality, and 
high availability of resources
would, 
possibly, not result in very high density and reproduction for super-AIs. 
Which is consistent with life history theory.\cite{ellis2009fundamental}

%

\subsection{On Detecting Super-AIs}
Searches for feasible projects that are more likely to not be 
subject to diminishing returns
might have higher chances of positive findings
than searches for, 
say,
Kardashev type III societies (see above).
%

Super-AIs may,
for evolutionary reasons,
try to post\-pone succumbing to
adverse cosmological 
changes.

Since super-AIs are super, it might be most likely that 
they will succumb to
cosmological heat death
(cf.\ \cite{mack20}).




When the expansion of the universe is being felt more acutely,
and anything outside your galaxy cluster will be beyond the event horizon,
it would, 
possibly,
be better to have 
fewer and larger clusters
rather than
more and smaller clusters,
all 
else
being 
equal.

It is unclear though what, if anything, a possible preference for larger clusters would mean --- 
maybe small clusters are on the whole just fine.
Maybe
we don't know enough 
about e.g.\ physics
to answer 
this question 
(or the full Fermi paradox).
But as a hypothetical 
example,
if we 
observe 
configurations 
--- at lower redshifts,
but not at
very high
ones ---
that in
the far future
will result in useful clusters,
and to a larger extent than what we 
would otherwise  
expect,
then 
perhaps
we
could
consider the possibility that
those observations could be 
signs
of super-AI actions.

Further,
if super-AIs will 
succumb to
heat death,
then
possibly 
they could try to reduce entropy waste,
e.g.\ maybe by affecting star or black hole formation.

There are tensions in 
cosmology at high 
signifi\-cance,\cite{perivolaropoulos2022challenges, abdalla2022cosmology, aluri23}
prominently the Hubble tension, i.e.\ 
the discrepancy in the Hubble constant, $H_0$, of
how fast the universe is presently
expanding,\cite{riess2020expansion, verde2019tensions, di2021realm, schoneberg2022h0,
abdalla2022cosmology, perivolaropoulos2022challenges}
although see \cite{freedman2025status}.
%
In particular there is
the question to what extent there is a local underdensity,
and,
because of increased gravitational pull,
to what extent 
the underdensity
could affect $H_0$
measurements,\cite{keenan2013evidence}\cite[sec.\,14.14]{di2021realm}%
\cite[sec.\,7.8.3]{abdalla2022cosmology}%
\cite{boh20, shanks2019gaia, haslbauer2020kbc, kenworthy2019local,
camarena2022void, Giani_2024, Huterer2024}
although
the more recent studies tend to show that the local underdensity 
alone is not enough to relieve 
the Hubble tension.
%
As an 
albeit
hypothetical, speculative, 
and 
unrealistic
but still more concrete example
then, 
it might be possible that
evolution rooted super-AI actions could cause 
underdensities,
that we could maybe 
decipher.
For example, a ``$10\,\sigma$'' underdensity 
(where the `$10\,\sigma$' relates to the likelihood of randomly occurring in 
standard $\Lambda$CDM)
might
explain the $H_0$ 
tension,\cite{haslbauer2020kbc}
and it might be possible that such an underdensity could 
stem from
super-AI actions.


Similarly but 
perhaps less unrealistic, 
there might be 
interference with star or black hole formation, that we could maybe detect.%

Albeit speculative, hypothetical, and unlikely,
a simpler 
cosmological
model
could possibly
fit data better, if taken together with 
specific and postulated, and ideally ``natural'', super-AI actions.

Or maybe there isn't anything
not subject to diminishing returns, secularly.

\nocite{mazurenko2023simultaneous, sec21, brout22, tully2019cosmicflows}
\subsection{Conclusion}

In conclusion, looking at it from the other direction, we have 
these 
observations and principles:
\pagebreak[2]
\begin{itemize}
  \itemsep=0pt
  \item The Copernican or mediocrity principle\footnote{With `Copernican'
    taken in a loose sense: 
    The observer selection effect notwithstanding, and all else equal,
    it is more likely that things are continuous, or ``normal'' or Gaussian,
    rather than discrete,
    or unique. In this sense, all else equal, it is perhaps still more likely that, 
    for example, there are more habitable planets than just Earth,
    that
    life isn't that rare,
    or that highly intelligent life isn't all that unique. (Cf.\ \cite{whit22}.)}
  \item The equilibrium principle
  \item The seemingly likely abundance of old societies, perhaps
  \item The observation that just a single society ought to be able to quickly
        engender an exploration and a colonization of e.g.\ the Milky Way
  \item The Fermi paradox
  \item No evidence of
  non-cooperative old societies or super-AIs:
  the Solar System hasn't been made into paper clips e.g.\
  \item Seemingly no Kardashev type III societies
\end{itemize}
This suggests 
the possibility
that all old societies and super-AIs behave similarly in these regards,
because of things they have in common, for example 
cooperative evolutionary pressure.
A look at these common things, an extrapolation of 
human progression, 
and diminishing
beneficial returns from material resources,
indicate
that cooperation is 
increasingly
adaptive
as wealth increases,
and the possibility
that
on the whole
there will be no incentive
to for instance colonize entire galaxies,
which 
could
explain the Fermi paradox.


\appendix





\section{Super-AI}
\phantomsection
\label{sec:appendixSuperAI}

\subsection{Super-AI Is Likely Feasible}
Artificial super-intelligence (super-AI, ASI) seems likely to  
be 
developed.\cite{grace2024thousands,owidaitimelines,gra18,zhang22,grue19,gra22,walsh18,muehl15}
(See e.g.\ 
Nielsen \cite{niel15}, 
or Goodfellow et al.\ \cite{goodfellow2016deep},
for an introduction to AI based on neural networks. 
Or see
Mitchell's
non-technical 
\cite{mitch19}.)

\enlargethispage{-0.9\baselineskip}

\subsection{Fitter Super-AIs}

That super-AIs can be evolutionary fitter than e.g.\ humans ought to be
more or less a tautology: 
super-AIs can use biological evolutionary solutions, but also any other solution. 
For example, for energy, super-AIs could mimic how biological entities extract energy,
or use any of the other more efficient or powerful ways to get energy that are
physically possible.

\subsection{On What Super-AIs Are Like}


\subsubsection{Neural vs.\ Symbolic AI}
The neural approach to AI is
characterized by
fast feedback, learning, and evolution,
featuring continuous functions;\cite{niel15}
the approach has had some recent success.\cite{alphaZero17,mitch19}
This is in contrast to the symbolic approach, which is based more on rules and discontinuous 
functions.\cite{mitch19}

With the neural approach to AI in a broad sense, every biological but also cultural evolutionary 
solution to a problem ought to potentially inform the AI solution,
to the extent that it is useful.
(Cf.\ e.g.\ Hassabis et al.\ \cite{hass17},
and Mitchell \cite{mitch19}.)
So with the evolutionary approach to morality, those moral solutions can inform AI solutions.
(Cf.\ Loosemore \cite{loose14}.)
\nocite{etz16}
(Cf.\ \cite{horibe2026emergence, vallinder2024cultural,
brookins2023playing, aher2023using, guo2023gpt, leng2023llm}, too; 
large language model
AIs 
generally
show cooperate behavior, reflecting the underlying training corpus
(in \cite{vallinder2024cultural} mainly Claude 3.5 Sonnet showed cooperate behavior, 
while GPT-4o failed to do so, and Gemini 1.5 Flash only showed weak cooperation).)


While,
according to some evolutionary view of morality at least, 
it would be adaptive for 
a 
super-AI 
to
be 
cooperative,
the AI solution to a cooperative evolutionary problem could be more discerning and 
complex than biological solutions (cf.\
\hyperlink{a_T}{\(a_\top\)} in
section~\ref{subsec:feasibility}, and \cite{andre23evolutionary}).
%
Another difference could potentially be that
longevity might contribute to increased cooperativeness for super-AIs;
see section \ref{subsubsec:predProj}, e.g.\ \cite{lie22,and22}.
\section{An Algorithm for Colonizing a Galaxy Quickly}
\label{sec:appendixColAlg}

We'll view the galaxy as a complete Euclidean graph when it's useful.

We can make a static plan for a colonization at the outset.

Let $S\coloneqq\{s_1,\dots,s_n\}$ be the sites to be colonized.
Let $s_0$ be the start site.
Let $p_i$ be the number of colonization ``probes'' that $s_i$ can make
without any significant strain ($p_i$ can be $0$ even from the outset;
on average they should be $>1$, and we expect them to be much larger).
Let $A\coloneqq\{s_0\}$ be the set of active sites that colonize.
Let $\Pi\coloneqq\emptyset$ be the set of colonization paths to active sites taken so far.
We'll also collect in $\Pi'\coloneqq\emptyset$ all colonization paths ending in a site $s_j$
such that $p_j=0$ already from the outset.

\enlargethispage{-0.9\baselineskip}

Algorithm \ref{alg:galaxyColonization} will then produce a colonization plan.

\begin{algorithm}
\caption{Galaxy Colonization}
\label{alg:galaxyColonization}
\begin{algorithmic}
   \While {$S \neq \emptyset$}
      \parState{Let $s_j \! \in \! S$ 
         be the yet uncolonized site with shortest
         path $\pi+(s_i,s_j)$ to $s_0$ using sub-path $\pi\!\in\!\Pi$, with $\pi$ ending
         in $s_i$. (So $s_i\!\in\!A$.) Then $s_i$ is to colonize $s_j$:}
      \State $p_i \leftarrow p_i-1$
      \If {$p_i = 0$}
         \State $A \leftarrow A-\{s_i\}$
         \State $\Pi \leftarrow \Pi-\{\pi\}$
      \EndIf
      \State $S \leftarrow S-\{s_j\}$
      \If {$p_j \neq 0$}
         \State $A \leftarrow A+\{s_j\}$
         \State $\Pi \leftarrow \Pi+\{\pi+(s_i,s_j)\}$
      \Else
         \State $\Pi' \leftarrow \Pi'+\{\pi+(s_i,s_j)\}$
      \EndIf
   \EndWhile
\end{algorithmic}
\end{algorithm}

Let $\Pi''$ be $\Pi'$ plus the paths in $\Pi$ ending with a site $s_i$ that hasn't colonized a 
single site. Then $\Pi''$ covers exactly the sites to be colonized, plus the start site $s_0$.


Since the
colonization
is
to happen in parallel, the number of colonizing sites, and in particular the number of available 
colonization probes, will increase exponentially --- 
or rather cubically or quadratically when physical space constraints are taken into 
account (see section 
\interfootnotelinepenalty=2000
\ref{para:NoExpCol}).\footnote{Another
 \label{fn:PNP}
 way to see the absurdity of actual exponential colonization
 (or reproduction)
 is to note that it would lead to $\mathbf{P}\!=\!\mathbf{NP}$, i.e.\ a polynomial time
 algorithm for solving e.g.\ the prototypical $\mathbf{NP}$-complete problem $\mathbf{SAT}$
 (boolean satisfiability):
 Split the $\mathbf{SAT}$ problem in 
 two
 by instantiating a variable. Send out the
 two
 subproblems. Repeat until all variables are instantiated, then solve each problem instance. 
 Report back the result to the parent node, all the way to the start node. If
 there can be an exponential colonization, then everything takes 
 only polynomial time.
 Corollary,
 if $\mathbf{P}\!\neq\!\mathbf{NP}$, then there cannot be an infinite number of 
 accessible dimensions.}
\interfootnotelinepenalty=1000
%

The galaxy colonization algorithm has been
implemented, and tested
under conditions intended to be Milky Way
\interfootnotelinepenalty=7900
like.\footnote{You can download the free and open-source software here:
\linebreak[2]
\url{https://github.com/DanielVallstrom/galaxyColonization}}
The tests indicate that the colonization plans are good. For example, the length of the 
colonization path to the site colonized last is very close to the straight line distance to that 
last site from the start site $s_0$, with the colonization path having less than $1\%$ extra 
length.\footnote{Regarding the details of the galaxy colonization algorithm, you can first sort
the sites with regard to the distance to the start site $s_0$. (This can be done in linear time 
with bucket
 sorting.)

\nopagebreak[4]
Second,
the straight line distance is a lower bound for the length of the colonization paths.
Hence, when looking for the next colonization, and if you go through the sites in order, whenever 
you have a straight line distance that is greater than or equal to the shortest colonization path 
found so far, you can abort that search.
(So you might also want to start searching for a site $s_j$ to colonize before looking for a site 
$s_i$ to colonize from.)
Since the lengths of the colonization paths often are very close to the straight line distances, 
this pruning
works well.

The set of active sites, $A$, is never actually used, as can be seen in 
algorithm \ref{alg:galaxyColonization}. $\Pi$ is the structure actually used.
$A$ is only there for illustrative purposes.}
\interfootnotelinepenalty=1000



Say that the speed of the probes is $0.1\, c$ (0.1 of the speed of light).
Then, with a start site near the center and within, say,
$5\times 10^4$ light-years of 
everything in
the Milky Way, it would take
$5\times 10^5$ years to colonize
the 
whole
galaxy,
assuming that the colonization of a site and
the creation of new probes take relatively 
little
time.
(We can conceive getting something weighing 2 grams up to $0.1\, c$.\cite{lubin16}
And it ought to be possible for a super-AI to weigh little.
(2 grams corresponds to roughly 1 mole so about $10^{23}$ atoms. 
The adult male human brain has around 86 billion neurons.\cite{aze09})
For deceleration, using stars near the target, see \cite{Heller_2017}.)

%
%
%


\section{Evolutionary Mismatch, Trophic Efficiency, and Negative-Cost Dietary Signaling: On How to Eat}
\label{sec:howToEat}

A basis for life is 
exploiting energy.
Evolutionary solutions for humans include 
tastes for fat and sugar, 
presumably.\cite{bres13}

The 
``10\% rule'', a rule of thumb, 
says that 
there is a 90\% 
energy loss between trophic 
food chain levels;\cite{wu2026global, natgeo10pct}
a recent 
synthesis of empirical 
studies found even lower 
efficiency, namely a 5.92\% average, 
with a terrestrial efficiency
of 
1.52\%.\cite{wu2026global}

The first agricultural revolution began after the last 
ice age, around 12,000 years ago;
this is typically too little time for genetic
adaptations to play 
out.\cite{martin2013plants, carrera2011western}
 
Paleolithic hunter-gatherers consumed plant resources 
to a large 
extent.\cite{moubtahij2024isotopic, chen2024stable, 
martin2013plants, carrera2011western} 
%
While an omnivore diet surely was adaptive given variable
access to food,\cite{leitzmann2014vegetarian} 
that doesn't mean that for example 
meat or dairy is 
useful
or healthy in an
environment with excess food. 
Dairy and lactase persistence are recent, there is little current supply of
wild meat, 
and the diet of farmed animals differs from the
diet of the Paleolithic animals.
Regardless,
vegan and vegetarian diets are
typically
likely
healthier today.\cite{cra21, Clark201906908, or13, Dinu22112017, mach15, 
boeing2012critical, DrawdownPlant,
mod20, bui2024planetary, tonstad2013vegetarian, tonstad2009type,
leitzmann2014vegetarian}
\nocite{boeing2012critical}




The evolutionary solutions with tastes for fat, sugar, and salt \cite{bres13} 
were for a bygone environment that didn't have a virtually limitless 
supply of fatty, sugary, and salty products readily available. 
These solutions promote overconsumption 
where fat, sugar, and salt are in abundance
(see section \ref{subsubsec:env}).




Due to, for example, concerns for animals or the environment,
vegetarians and vegans
might ostensibly
signal considerateness and cooperativeness;%
\cite{sing22, sing21, gha22, jan16, hack22, stoll17, kalte21, mod20, 
dhont2024health, coel19,
hop20, sparkman2017dynamic, ruby12, rosenfeld2018psychology, 
leitzmann2014vegetarian}
typical meat production 
harms the 
environment \cite{Clark201906908, stoll17, mach15, stein06, kraham17, DrawdownPlant,
cra21, mod20, ritchie2024not, bui2024planetary,
willett2019food, scarborough2023vegans}. 
%
%
For the prevalence of veganism and vegetarianism, see
\cite{pfeiler2018examining, ruby12, 
rosenfeld2018psychology, paslakis2020prevalence, chen2025we, ruby2016attitudes, 
WWF17, owidvegan, sim18, fran20}.
\nocite{atari20}

\protect\interfootnotelinepenalty=9100
The possible signaling by vegans
and vegetarians
should have negative cost since plant resources are one or two
orders of magnitude more efficient and cheaper,
and are likely healthier.
(For examples of specific costs, 
see \cite{OWiDMeatCost, owidlanduse}\cite[\hspace{-0.15em}ch.\,5]{ritchie2024not}.)
\nocite{eshel2025us}
\nocite{poore2018reducing}
\protect\nocite{davis18, vos14, pimm14, new18, ipbes19, kol14, owidbiodiversity}


\protect\section{Evolutionary Mismatches and Signals with Negative Cost}
\protect\label{sec:mismatchNegCost}

The above discussion (appx.\ \ref{sec:howToEat}) suggests that 
honest signals of
veganism,
and to a lesser extent
vegetarianism, are examples of signals with negative cost. 
Cf.\ \cite{sza23, sza22}\protect\footnote{\hspace{-0.17em}\hyperlink{sigSimRef}{\(^{\uparrow}\)}%
\hspace{0.23em}Here
is a free and open-source program
\label{fn:sigSim}
that simulates the development of signaling, 
showing
that negative cost signals can evolve,
following 
\cite{sza22}:
\url{https://github.com/DanielVallstrom/signalSim}}.
%
The idea that the
absolute 
cost is 
of the essence for
signals is a misunderstanding;
if a signal happens to have negative cost,
all the better; what matters are differences.\cite{sza23, sza22}
%
For example, nasal breathing during 
sleep might signal fitness,
and 
might 
have negative cost.
%
(`Signal' is used broadly throughout and encompasses cues and behaviors.)


\subsection{Emotional Mismatches}

Other acts
similar 
to veganism,
signaling cooperativeness, 
or other fitness benefits,
and with negative cost, might include:
running or cycling instead of driving; 
not smoking;
not habitually using e.g.\ alcohol or other drugs, pornography, social media;
forgoing habitual gambling;
forgoing 
counterproductive
child-rearing
practices, e.g.\ 
excessive pampering,
overfeeding;
non-consumerism.

\subsection{Cultural Mismatches}

Another example of 
an act
signaling cooperativeness
with negative cost 
might be 
atheism, instead of religiousness with costly practices,
where atheism correlates with cooperativeness (cf.\ \cite[esp.\,ch.\,4]{fitouchi2025prosocial}).
For instance, 
atheists 
in various countries, e.g.\ 
the USA,
express values,
e.g.\ 
higher levels of 
trust and emancipative values,
that might correlate with increased cooperativeness 
\cite{darcy2026trust, knack1997does, algan2014trust}, 
compared to religious people in the same country
\cite[ch.\,4]{fitouchi2025prosocial}\cite{WVS}.
\hypertarget{emancFigRef}{See}
figures \ref{fig:relAndTrust} 
and
\ref{fig:emanc}.

\begin{figure}
  \centering

  \begin{tikzpicture}

    \begin{axis}[
      ybar,
      legend style={at={(-0.22,1.03)},
        anchor=south west},
      symbolic x coords={``religious'',``non-religious'',``atheist''},
      xtick=data,
      legend columns=5,
      legend style={font=\scriptsize},
      width = 1.00\linewidth,
      height = 0.81\linewidth,
      bar width= 4pt,
      ybar=1pt,
      yticklabel={$\pgfmathprintnumber{\tick}\%$},
      grid style = {very thin, dashed, gray!40},
      minor y tick num=3,
      ymajorgrids=true,
      yminorgrids=true,
      xmajorgrids=false,
      xminorgrids=false,
      enlarge x limits={abs=1.2cm},
      extra y ticks={10,30,50,70,90},
      ymin=0, ymax=90,
      ylabel = {share saying ``most can be trusted''}
    ]


    \addplot +[pattern=north west lines,
               error bars/.cd, y dir=both, y explicit] coordinates {
      (``religious'',23.6339) +- (``religious'',3.0776)
      (``non-religious'',28.9474) +- (``non-religious'',3.4973)
      (``atheist'',35.2113) +- (``atheist'',4.5357)
    };
    \addlegendentry{FRA 2018}

    \addplot +[error bars/.cd, y dir=both, y explicit] coordinates {
      (``religious'',34.955) +- (``religious'',2.8051)
      (``non-religious'',32.0783) +- (``non-religious'',3.5504)
      (``atheist'',38.5027) +- (``atheist'',6.9744)
    };
    \addlegendentry{SGP 2020}

    \addplot +[pattern=dots,
               error bars/.cd, y dir=both, y explicit] coordinates {
      (``religious'',23.8372)  +- (``religious'',0.2724)
      (``non-religious'',35.0329) +- (``non-religious'',0.4535)
      (``atheist'',44.1981) +- (``atheist'',0.8556)
    };
    \addlegendentry{World}
    
    \addplot +[error bars/.cd, y dir=both, y explicit] coordinates {
      (``religious'',38.4669)  +- (``religious'',2.5287)
      (``non-religious'',39.4475) +- (``non-religious'',3.1843)
      (``atheist'',48.9796) +- (``atheist'',6.2597)
    };
    \addlegendentry{USA 2017}

    \addplot +[error bars/.cd, y dir=both, y explicit] coordinates {
      (``religious'',39.7631)  +- (``religious'',3.9458)
      (``non-religious'',40.0000) +- (``non-religious'',4.7134)
      (``atheist'',48.7179) +- (``atheist'',7.8437)
    };
    \addlegendentry{ES 2017}

    \addplot +[error bars/.cd, y dir=both, y explicit] coordinates {
      (``religious'',42.819)  +- (``religious'',2.5066)
      (``non-religious'',43.6585) +- (``non-religious'',2.1470)
      (``atheist'',51.3369) +- (``atheist'',3.5820)
    };
    \addlegendentry{GB 2020}

    \addplot +[error bars/.cd, y dir=both, y explicit] coordinates {
      (``religious'',47.2262)  +- (``religious'',2.6095)
      (``non-religious'',49.2754) +- (``non-religious'',2.3135)
      (``atheist'',53.912) +- (``atheist'',3.4160)
    };
    \addlegendentry{CAN 2020}

    \addplot +[pattern=north east lines,
               error bars/.cd, y dir=both, y explicit] coordinates {
      (``religious'',59.3241)  +- (``religious'',2.3653)
      (``non-religious'',64.2424) +- (``non-religious'',2.2050)
      (``atheist'',68.7151) +- (``atheist'',3.9216)
    };
    \addlegendentry{NL 2022}

    \addplot +[error bars/.cd, y dir=both, y explicit] coordinates {
      (``religious'',57.8462)  +- (``religious'',5.3687)
      (``non-religious'',59.4828) +- (``non-religious'',4.4670)
      (``atheist'',68.9922) +- (``atheist'',7.9817)
    };
    \addlegendentry{NZ 2020}

    \addplot +[pattern=crosshatch,
               error bars/.cd, y dir=both, y explicit] coordinates {
      (``religious'',74.2243)  +- (``religious'',4.1882)
      (``non-religious'',74.1768) +- (``non-religious'',3.5711)
      (``atheist'',81.7308) +- (``atheist'',7.4266)
    };
    \addlegendentry{NOR 2018}

    \end{axis}
  \end{tikzpicture}

  \vspace{-0.55em}
  \caption[Share saying that most people can be trusted.]
          {Share saying that most people can be trusted, for each
           of the three disjoint groups. 
           Data from IVS (EVS and WVS),\cite{ivs2022} with non-answers discarded.
           'World' means all 92 countries included in the 2017-2022 survey.
           95\% confidence intervals.\footnotemark{}}
          
  \label{fig:relAndTrust}

  \vspace{-0.10em}
\end{figure}

\if 01
Run e.g.
./ivsExtract  -f EVS_WVS_Joint_Csv_v5_0.csv -F 6 -c 0
for world data.
And e.g.
./ivsExtract  -f EVS_WVS_Joint_Csv_v5_0.csv -F 6 -o ivs2017_atheism_trust.result
for country data.
Newer versions of ivsExtract will output CIs too.

USA
2017
2596 total
ignoring non-answers:
1425-0    = 1425 ``religious'' - 3 non-answers = 1422
  547 -0    = 547 ``most can be trusted''   0.384669  38%
  1422-547  = 875 need to be very careful  62%
  1425-1422 = 3   no answer
 
1670-1425 = 245  ``atheist''
  1545-1425 = 120 most can be trusted  0.489796  49%
  1670-1545 = 125 need to be very careful  51%
  0 no answer

1690-1670 = 20   non-answers
  4 most can be trusted
  11 need to be very careful
  5 no answer

2596-1690 = 906  ``non-religious'' - 1 = 905
  2047-1690 = 357 most can be trusted  0.394475  39%
  2595-2047 = 548 need to be very careful  61%
  1 no answer
  
The above data are the same as the data below extracted using ivsExtract:
 x y: counter
 0 0: 5
 0 1: 4
 0 2: 11
 1 0: 3
 1 1: 547
 1 2: 875
 2 0: 1
 2 1: 357
 2 2: 548
 3 0: 0
 3 1: 120
 3 2: 125

------

world
Results with non-answers discarded:
 x y: count  share  count / total
 1 1: 22401  0.238372  22401 / 93975
 1 2: 71574  0.761628  71574 / 93975
 2 1: 14896  0.350329  14896 / 42520
 2 2: 27624  0.649671  27624 / 42520
 3 1:  5721  0.441981   5721 / 12944
 3 2:  7223  0.558019   7223 / 12944
 
 x y: count  share     count / total   CI
 1 1: 22401  0.238372  22401 / 93975  0.002724
 1 2: 71574  0.761628  71574 / 93975  0.002724
 2 1: 14896  0.350329  14896 / 42520  0.004535
 2 2: 27624  0.649671  27624 / 42520  0.004535
 3 1:  5721  0.441981   5721 / 12944  0.008556

CAN
 x y: count  share  count / total
 1 1:   664  0.472262    664 /  1406
 1 2:   742  0.527738    742 /  1406
 2 1:   884  0.492754    884 /  1794
 2 2:   910  0.507246    910 /  1794
 3 1:   441  0.53912    441 /   818
 3 2:   377  0.46088    377 /   818
 
Great Britain
 x y: count  share  count / total
 1 1:   641  0.428    641 /  1497
 1 2:   856  0.572    856 /  1497
 2 1:   895  0.437    895 /  2050
 2 2:  1155  0.563   1155 /  2050
 3 1:   384  0.513    384 /   748
 3 2:   364  0.487    364 /   748 
 x y: count  share     count / total   CI
 1 1:   641  0.428190    641 / 1497  0.025066
 1 2:   856  0.571810    856 / 1497  0.025066
 2 1:   895  0.436585    895 / 2050  0.021470
 2 2:  1155  0.563415   1155 / 2050  0.021470
 3 1:   384  0.513369    384 /  748  0.035820
 3 2:   364  0.486631    364 /  748  0.035820
 
Netherlands
Results with non-answers discarded:
 x y: count  share     count / total   CI
 1 1:   983  0.593241    983 / 1657  0.023653
 1 2:   674  0.406759    674 / 1657  0.023653
 2 1:  1166  0.642424   1166 / 1815  0.022050
 2 2:   649  0.357576    649 / 1815  0.022050
 3 1:   369  0.687151    369 /  537  0.039216
 3 2:   168  0.312849    168 /  537  0.039216
 
NZ
Results with non-answers discarded:
 x y: count  share  count / total
 1 1:   188  0.578462    188 /   325
 1 2:   137  0.421538    137 /   325
 2 1:   276  0.594828    276 /   464
 2 2:   188  0.405172    188 /   464
 3 1:    89  0.689922     89 /   129
 3 2:    40  0.310078     40 /   129
 
NOR 
Results with non-answers discarded:
 x y: count  share  count / total
 1 1:   311  0.742243    311 /   419
 1 2:   108  0.257757    108 /   419
 2 1:   428  0.741768    428 /   577
 2 2:   149  0.258232    149 /   577
 3 1:    85  0.817308     85 /   104
 3 2:    19  0.182692     19 /   104
 
Singapore 
Results with non-answers discarded:
 x y: count  share  count / total
 1 1:   388  0.34955    388 /  1110
 1 2:   722  0.65045    722 /  1110
 2 1:   213  0.320783    213 /   664
 2 2:   451  0.679217    451 /   664
 3 1:    72  0.385027     72 /   187
 3 2:   115  0.614973    115 /   187

country: 724  724 	Spain 	 
rows with country this round: 1209
Results with non-answers discarded:
 x y: count  share     count / total   CI
 1 1:   235  0.397631    235 /  591  0.039458
 1 2:   356  0.602369    356 /  591  0.039458
 2 1:   166  0.400000    166 /  415  0.047134
 2 2:   249  0.600000    249 /  415  0.047134
 3 1:    76  0.487179     76 /  156  0.078437
 3 2:    80  0.512821     80 /  156  0.078437

France
Results with non-answers discarded:
 x y: count  share  count / total
 1 1:   173  0.236339    173 /   732
 1 2:   559  0.763661    559 /   732
 2 1:   187  0.289474    187 /   646
 2 2:   459  0.710526    459 /   646
 3 1:   150  0.352113    150 /   426
 3 2:   276  0.647887    276 /   426
 
Czechia, CZE
Results with non-answers discarded:
 x y: count  share  count / total
 1 1:   220  0.238871    220 /   921
 1 2:   701  0.761129    701 /   921
 2 1:   425  0.315048    425 /  1349
 2 2:   924  0.684952    924 /  1349
 3 1:   149  0.321814    149 /   463
 3 2:   314  0.678186    314 /   463
 
China
country: 156 
rows with country: 3036
Results with non-answers discarded:
 x y: count  share  count / total
 1 1:   286  0.606    286 /   472
 1 2:   186  0.394    186 /   472
 2 1:   957  0.653    957 /  1466
 2 2:   509  0.347    509 /  1466
 3 1:   699  0.675    699 /  1036
 3 2:   337  0.325    337 /  1036
  
SWE 
2017
  
1194 total

(FJ) (BZ)

347 religious (1) - 7 non-answers = 340
  233 most can be trusted (1)   0.685294 
  107 careful (2)  
  7 other
  
605 non-religious (2) - 9 non-answers = 596
  385 trust    0.645973
  211 careful
  9 other
  
212 atheist (3) - 2 non-answers = 210
  151 trust    0.719048
  59 careful
  2 other
  
\fi

%
%
\footnotetext{The
\label{fn:ivsExt}
           data were extracted using the free and open-source 
           program ivs\-Extract, available here:
           \url{https://github.com/DanielVallstrom/ivsExtract}}          
%
%

Other 
examples of negative cost acts 
signaling cooperativeness 
might include:
not practicing
footbinding;%
\footnote{The spread 
\label{fn:matriPatri}
of matriliny and patriliny depends on the 
benefit of resources for daughters ($B_D$), compared to
the benefit for sons and 
paternity probability 
($B_S\times P$).\cite{holden2003spread}
%
Rising relative benefit for daughters will affect 
practices.}
not practicing
female genital mutilation;
%
%
%
%
forgoing manicured gardens and lawns;
forgoing overt romantic signals \cite{dubour26paradox};
and perhaps bows instead of handshakes.
%
%
%
%
%

\subsection{General Principle}

Entities signal fitness benefits.
The absolute cost of the signal doesn't matter, ultimately.
What matters is that a signal correlates with a type.
Above, for instance, a type that cannot do nasal breathing during
sleep cannot do it
with negative cost.
Proximately, signals might be costly, of course
(although, for example, even peacock plumage is \emph{not} very
costly \cite{sza23, sza22}).
And everything is situated;
the same signal may correlate with cooperativeness
in one environment, while the opposite 
is
true
in another environment. 

More generally then, 
high types
can forgo 
signals correlated with 
lower types,
if 
the high types
differentiate 
themselves nevertheless.
In particular, everything maladaptive could be forgone,
if others agree on the maladaptiveness.

\subsection{Solutions}
\label{sec:solutions}

When 
our environment changes rapidly,
as 
in our exponential increase in available resources,
and when 
for example
our 
impulses
and emotions cannot keep up with that pace,
laws, norms, habits 
--- cultural evolution ---
must compensate
(see sec.\,\ref{subsubsec:env}). 
Cf.\ ``evolutionary mismatch'',
where adaptations become maladaptive because of environmental changes; 
see e.g.\ \cite{li2018evolutionary, tooby1990past}.
Cf.\ too the Edward O Wilson quote 
  ``The real problem of humanity is the following: we have Paleolithic emotions,
    medieval institutions, and godlike technology''.\cite{wilson2009problemQuote}

In particular, if the negative consequences of
prohibiting a maladaptive, emotional environment
are only reduced available resources that show
diminishing returns, then
those costs
are inconsequential.




Assortment 
can
facilitate the invasion of negative cost signals.
As in the footbinding case where 
families came to an agreement 
beforehand.\cite{mackie1996ending}

\nocite{pazhoohi2017religious}

\nocite{centola2018experimental}

%

%

More generally, 
%
we can, as individuals, 
endeavor to
avoid 
maladaptive
environments,
especially
addictive or habitual ones.\footnote{Individual 
strategies might include intention implementations (if-then plans, 
implementation intentions, 
where you simulate a habit beforehand; see the bibliography for 
sec.\ \ref{subsubsec:env}, e.g.\ \cite{gollwitzer2006implementation, 
gollwitzer1999implementation, sheeran2016intention, sheeran2002intention}).
See also \cite{buabang2025leveraging}.}
%
%
We can, as a society, limit or prohibit maladaptive environments in the first place.
And we can 
endeavor to
create environments that don't incentivize entities, e.g.\ companies,
to in turn create 
maladaptive environments.
%
%
%
%




\nocite{albarracin2024determinants}

\subsection{Mismatched Aversive Sensations}


There are also 
evolved 
aversive
physiological
sensations
that 
no longer reliably
indicate
something negative.
Sensations that evolved
to save energy could now more often beneficially be ignored.
%
For instance, 
our physical and cognitive fatigue systems evolved in an environment where 
saving energy was broadly
adaptive.
Hence, saving energy in general, when opportune, might have been adaptive.
Or
%
having a continuous system, instead of a binary one with abrupt cessations, 
might have been adaptive.
However, we now have more energy available, and more control, and e.g.\ taking 
an elevator 
one floor
might be maladaptive;
we could 
ignore maladaptive discomfort, with negative 
cost.\footnote{Another
explanation of habits and addictions could be that
a proven heuristic is statistically safer than
repeated 
conscious
deliberation,
with an exponential probability of an eventual bad outcome, compared to the guaranteed 
habitual outcome (with the
eventual probability of the bad outcome being $1 - (1-b)^{n}$, 
if $b$ is the probability
of the bad outcome for one conscious decision).
Or simply, rather, that the habitual outcomes are, or were, on average better than deliberate outcomes.
However, again, that was in a bygone environment;
we could now increasingly reassess and break 
maladaptive
habits and addictions.}
%
Cf.\ 
\cite{pessiglione2025origins,
thornton2024athletes, tesarz2012pain, geva2013enhanced, pettersen2020pain,
jamadar2024metabolic}.%
\nocite{kool2010decision}%
\nocite{shenhav2017toward}%
%
%
\nocite{marcora2009perception}%
\nocite{behrens2023fatigue}%

\nocite{courtwright2019age}


\subsection{Discussion}
\label{sec:discussion}

Tooby and Cosmides argued that  
``improvements in the technologies of food production and
in public health and medicine are two areas of `behavior' that have vastly
decreased prereproductive mortality and increased lifetime reproductive
success. They are largely --- perhaps entirely --- responsible for the population
increase''.\cite{tooby1990past}
When it comes to present or future impacts, not even that much might be true:
much of increased food supply will increasingly have negative effects;
even most of the gains in medicine for 
``prereproductive mortality and increased lifetime reproductive success''
might
be behind us.

However, 
if so much is 
disadvantageous, then
information accumulation, cultural evolution, cultural technologies
(that enabled 
the advancements in food and health
in the first place),
to handle 
all the 
evolutionary mismatches,
should be 
advantageous.
For example, 
reining in 
companies, 
individuals,
or states 
that 
create or exploit
maladaptive
environments
should be beneficial.

\section{On the Formal Definition of Diminishing Returns}
\label{appx:DRAlt}

An alternative to definition \ref{def:DR} of 
diminishing returns would be 
\begin{defn}[$DR'$]
\label{def:DRAlt}
$f \not\in DR'  \Leftrightarrow
\exists n\, x^{1/n} \in O(f)$
\end{defn}
\noindent
That is, 
$f \!\in\! DR' \Leftrightarrow \forall n\, x^{1/n} \!\not\in\! O(f)$,
with $DR\subseteq DR'$.
\begin{theorem}
\label{theorem:DRAltClosed}
If $f\in DR'$, and $g\in O(x^k)$, 
then $g\circ f \in DR'$.
\end{theorem}
\vspace{-1.0em}
\begin{proof}
Assume that $f\in DR'$, and $g\in O(x^k)$.
Suppose that $g\circ f \not\in DR'$. 
Then there is an $n$ such that $x^{1/n}\in O(g\circ f)$.
But then 
$x^{1/(kn)}\in O(f)$,
a contradiction.
\end{proof}

For the other direction,
assume that $f$ doesn't have 
diminishing returns according to definition \ref{def:DRAlt}.
Then there is an $n$ such that
$x^{1/n} \in O(f)$. 
But then
any
decreasing returns of $f$ can be overcome by a feasible amount of extra work,
by duplicating the result of $f$ 
$x^n$ times,
with the result 
$x^n\circ f$,
for, say, an above unit 
marginal return.
Hence, $f$ 
has
non-diminishing returns also in some intuitive sense.

However, certain, although perhaps unnatural, edge cases would, with
definition \ref{def:DRAlt}, not be what you maybe 
want. For example, $\lvert\sin x\rvert x$ doesn't have diminishing returns,
and $\lvert\sin x\rvert x \not\in DR$, but 
$\lvert\sin x\rvert x \in DR'$. See figure \ref{fig:DRAlt}.

\begin{figure}
  \centering

  \begin{tikzpicture}

    \begin{axis}[
      legend style = {at={(0.37,0.97)}, anchor=north west},
      width = 0.99\linewidth,
      height = 0.99\linewidth,
      axis y line = left,
      axis x line = bottom,
      axis line style = {very thick},
      grid style = {thin, densely dotted, black!50},
      xmin=0, xmax=17,
      ymin=0, ymax=17,
      xlabel = $x$,
      ylabel = {\(f(x)\)}]

      \addplot[smooth, domain=0:17, samples=500, very thick]{x*abs(sin(deg(x)))};
      \addlegendentry{\(\lvert\sin x\rvert x\)}




      \addplot[smooth, domain=1:3, samples=100, thin]{(1 - abs(x-2)) * 2};
      \addlegendentry{\(\mathfrak{d}'x\)}

      \addplot[smooth, domain=1.75:2.25, samples=100, dashed]{4*(1/4 - abs(x-2)) * 2*2};
      \addlegendentry{\(\mathfrak{d}''x\)}


      \addplot[smooth, domain=3:5, samples=100, thin]{(1 - abs(x-4)) * 4};

      \addplot[smooth, domain=7:9, samples=100, thin]{(1 - abs(x-8)) * 8};

      \addplot[smooth, domain=15:17, samples=100, thin]{(1 - abs(x-16)) * 16};


      \addplot[dashed] coordinates {(4,0)(4,17)};
      \addplot[dashed] coordinates {(4,16.95)(4,0)};
      
      \addplot[dashed] coordinates {(16,0)(16,17)};
      \addplot[dashed] coordinates {(16,16.95)(16,0)};
            
      
      

    \end{axis}
  \end{tikzpicture}

  \vspace{-0.55em}
  \caption[|sin x|x is in 
           DR'.]{$\lvert\sin x\rvert x \not\in DR$, but $\lvert\sin x\rvert x \in DR'$.
                 $\mathfrak{d}'\in DR'$, but $\mathfrak{d}'\not\in DR$.
                 $\lvert\sin x\rvert x \not\in DR''$ and 
                 $\mathfrak{d}'\in DR''$.
                 $\mathfrak{d}'' \in O^{\smallint}(1)$,
                 but $\mathfrak{d}'' \not\in O(2^x)$.}
  \label{fig:DRAlt}
  \vspace{-0.10em}
\end{figure}

On the other hand, there are continuous functions with
diminishing returns, that are in $DR'$, but not in $DR$.
For instance,
let $\mathfrak{d}'(x) = 0$, except for 
$2^n - 1 \leq x \leq 2^n + 1$,
$n\in\mathbb{N}^+$,
where $\mathfrak{d}'(x) = 
(1 - \lvert x-2^n \rvert) \cdot 2^n$.
Then $\mathfrak{d}'$ has diminishing returns, and
$\mathfrak{d}'\in DR'$, but $\mathfrak{d}'\not\in DR$.
See figure~\ref{fig:DRAlt}.
To remedy this one could consider integrals.
We'll replace the simple function value comparison in the definition of $O$
with a comparison of integrals over some minimal or larger interval:
\begin{defn}[$O^{\smallint}$, integral $O$]
\label{def:O'}
$f \in O^{\smallint}(g)  \Leftrightarrow$
\[\exists c\exists x_0 \, \forall x\!\!\geq\!\!x_0 \, 
\exists y_0\!\!>\!\!x\, \forall y\!\!\geq\!\!y_0 
\int_{x}^{y}\!\!f(z) \,dz \,\leq\, c\! \int_{x}^{y}\!\!g(z) \,dz\]
\end{defn}
\begin{defn}[$DR''$]
\label{def:DR''}
$f \in DR'' \Leftrightarrow \forall n\, f \in O^{\smallint}(x^{1/n})$
\end{defn}
\begin{theorem}
\label{theorem:DR''Closed}
If $f\in DR''$, and $g\in O(x^k)$, 
then $g\circ f \in DR''$.
\end{theorem}
\vspace{-1.0em}
\begin{proof}
Assume that $f\in DR''$, and $g\in O(x^k)$.
Let $n\in \mathbb{N}$.
But $f\in O^{\smallint}(x^{1/(kn)})$. Hence $g\circ f \in O^{\smallint}(x^{1/n})$.
\end{proof}

%
\vspace{-0.1em}
Let $\prescript{n}{}{a}$ be a tetration tower $a^{a^{\iddots^a}}$ 
with $n-1$ exponentiations,
with $\prescript{1}{}{a} = a$ having height $1$ and
$\prescript{n}{}{a}$ having height $n$.
Let $\mathfrak{d}''(x) = 0$, except for 
$\prescript{n}{}{2} - 1/(\prescript{\prescript{n}{}{2}}{}{2}) \leq x \leq 
 \prescript{n}{}{2} + 1/(\prescript{\prescript{n}{}{2}}{}{2})$,
$n\in\mathbb{N}^+$,
where $\mathfrak{d}''(x) = 
(\prescript{\prescript{n}{}{2}}{}{2})
(1/(\prescript{\prescript{n}{}{2}}{}{2}) - 
\lvert x-\prescript{n}{}{2} \rvert) \cdot 
\prescript{\prescript{n}{}{2}}{}{2}$.
Then $\mathfrak{d}''$ is continuous,
$\mathfrak{d}'' \not\in O(2^x)$, 
but $\mathfrak{d}'' \in O^{\smallint}(1)$.
%
%
See figure~\ref{fig:DRAlt}.
($\prescript{\prescript{n}{}{2}}{}{2}$
could
be replaced by any function,
and the ``spikes'' could be centered at other intervals,
e.g.\ at every $n$ instead of at $\prescript{n}{}{2}$ as here.)

Cf.\ computational average-case complexity.

\subsection{A Theory of Production}

Algorithms or programs can be viewed as producing answers.
That algorithms solving an $\mathbf{NP}$-complete problem
have diminishing returns is equivalent to
$\mathbf{P}\!\neq\!\mathbf{NP}$.
It may be fruitful
to try to develop 
a more complete and grounded
theory of
production,
extending beyond the 
dual notions of 
diminishing returns and infeasibility.

\section{Modeling the Evolution of Cooperation and Fairness Under Diminishing Returns}
\label{sec:evoMod}

Andr{\'e} and 
Baumard have modeled how fairness evolves.\cite{andre11}
Here we will attempt
something similar but 
with the added 
assumption of some diminishing returns.

A round in the 
attempted
model starts off with an initial dictator or ultimatum type game. Typically
the fixed proposers are less than or equal to the responders, 
which creates pressure on the proposers to offer the least 
meaningful
amount possible.\cite{andre11}
Still,
the gains from this initial cycle are assumed to be under diminishing returns.
%

After the initial game, a second cycle of cooperation begins. This second cycle is some type
of dictator or ultimatum game too. The top cooperator is paired with the top
lower half cooperator, with the top cooperator as proposer, and the top lower half cooperator 
as responder. The second-top cooperator is paired with the second-top lower half cooperator.
And so on.

There are 
variations of the model: The cooperation tendencies between
the two cycles can be separate or correlated. The cooperation tendencies of the responders may
directly influence the immediate gains of the proposers in the second cycle, to some extent.
To break ties in the second cycle of cooperation, one can treat the cooperation tendencies in
the first cycle as signals (c.f.\ \cite[\S\,2.2, p.\,7]{and22})
and favor entities with higher first cycle offers, when applicable.

The model and its variations have been 
implemented.\footnote{The free and open-source software is available at:
\linebreak[2]
\url{https://github.com/DanielVallstrom/dimRetEvoSim}}
Each round takes linear time in the size of the population.

Simulation runs suggest, as 
is to be 
expected, that the first cycle of cooperation is of
secondary importance. 
For example,
offers less than half of the pie will be roughly equally good
when it comes to direct, immediate gains. 
Hence, if there are other aspects to the offer, e.g.\ ``signaling'' or that gains in
the second cycle of cooperation are affected, those other aspects will 
come into play;
offers will be larger under diminishing returns than, say, linear returns.

Simulation runs suggest too that cooperation and fairness evolve. 
In particular,
if the responders 
contribute 
to the gain of the proposers in the second cycle of cooperation,
then the proposers eventually offer 
more than half the pie,
because of pressure
to be paired with 
good
responders.
If the offer in the first cycle cannot be much lower than the second cycle offer, 
then the first cycle offer too will eventually be more than half the pie.
If there is no extra contribution from the responders in the second cycle,
then the offers will settle around half the pie.

\if 01
\section{Bibliography}
\fi


\renewcommand{\refname}{Bibliography\protect\footnote{Click or hover note icons for abstracts, if your viewer supports it.}}
%
%
{\small
\bibliography{morality_adaptations}}


\FloatBarrier

\vspace{2mm}

\section*{Supplementary Figures}

\setcounter{figure}{0}
\makeatletter
\renewcommand{\thefigure}{S\arabic{figure}}

\vspace{1mm}

\begin{figure}[!htb]
  \centering

  \begin{tikzpicture}

    \begin{axis}[
      legend style={font=\footnotesize},
      legend style={fill=none},  
      legend cell align=left,
      width = 0.88\linewidth,
      height = 0.75\linewidth,
      axis y line = left,
      axis x line = bottom,
      axis line style = {very thick},
      xticklabel style={/pgf/number format/1000 sep={}},
      yticklabel={$\pgfmathprintnumber{\tick}\%$},
      legend pos = north west,
      grid style = {very thin, dashed, gray!40},
      grid = major,
      xmin=1958, xmax=2021,
      ymin=0, ymax=100,
      xlabel = {year},
      ylabel = {approve of interracial marriage}]

      \addplot+ [mark size=0.3] 
        table[x=year,y=approval,col sep=comma]
             {interracialMarriageApprovalUS.csv}; 
      \addlegendentry{USA}

    \end{axis}
  \end{tikzpicture}

  \vspace{-0.55em}
  \caption[Share of people saying that they approve of interracial marriage, 1958 to 2021.]
          {\hspace{-0.65em}\hyperlink{raceFigRef}{\(^{\uparrow}\)}\hspace{0.66em}Share
           of people saying that they approve of interracial marriage, 1958 to 2021,
           USA. Gallup. See \cite{mccarthy2021us}.
           (Valeat quantum.)
          }
          
  \label{fig:race}

  \vspace{1mm}

\end{figure}

\if 01

country: 203  203 	Czechia 	Before 1993: part of Czechoslovakia 
wave: 2
rows with country this round: 924
Results with non-answers discarded:
 x y: count  share     count / total
 1 1:   109  0.333333    109 /  327
 1 2:   218  0.666667    218 /  327
 2 1:   140  0.283976    140 /  493
 2 2:   353  0.716024    353 /  493
 3 1:    30  0.294118     30 /  102
 3 2:    72  0.705882     72 /  102

country: 203  203 	Czechia 	Before 1993: part of Czechoslovakia 
wave: 3
rows with country this round: 1147
Results with non-answers discarded:
 x y: count  share     count / total
 1 1:   137  0.297826    137 /  460
 1 2:   323  0.702174    323 /  460
 2 1:   136  0.283333    136 /  480
 2 2:   344  0.716667    344 /  480
 3 1:    28  0.250000     28 /  112
 3 2:    84  0.750000     84 /  112

country: 203  203 	Czechia 	Before 1993: part of Czechoslovakia 
wave: 7
rows with country this round: 1200
Results with non-answers discarded:
 x y: count  share     count / total
 1 1:    87  0.273585     87 /  318
 1 2:   231  0.726415    231 /  318
 2 1:   236  0.406196    236 /  581
 2 2:   345  0.593804    345 /  581
 3 1:   107  0.421260    107 /  254
 3 2:   147  0.578740    147 /  254
 
 Czechia, CZE
Results with non-answers discarded:
 x y: count  share  count / total
 1 1:   220  0.238871    220 /   921
 1 2:   701  0.761129    701 /   921
 2 1:   425  0.315048    425 /  1349
 2 2:   924  0.684952    924 /  1349
 3 1:   149  0.321814    149 /   463
 3 2:   314  0.678186    314 /   463

country: 250  250 	France 	 
wave: 5
rows with country this round: 1001
Results with non-answers discarded:
 x y: count  share     count / total
 1 1:    88  0.191304     88 /  460
 1 2:   372  0.808696    372 /  460
 2 1:    54  0.151685     54 /  356
 2 2:   302  0.848315    302 /  356
 3 1:    42  0.250000     42 /  168
 3 2:   126  0.750000    126 /  168

France
Results with non-answers discarded:
 x y: count  share  count / total
 1 1:   173  0.236339    173 /   732
 1 2:   559  0.763661    559 /   732
 2 1:   187  0.289474    187 /   646
 2 2:   459  0.710526    459 /   646
 3 1:   150  0.352113    150 /   426
 3 2:   276  0.647887    276 /   426
 
\fi

\begin{figure}[!htb]
  \centering

  \begin{tikzpicture}

    \begin{axis}[
      legend style={font=\footnotesize},
      legend style={fill=none},  
      legend cell align=left,
      width = 0.83\linewidth,
      height = 1.1\linewidth,
      axis y line = left,
      axis x line = bottom,
      axis line style = {very thick},
      xticklabel style={/pgf/number format/1000 sep={}},
      legend style={at={(axis cs:2022,-0.15)},anchor=south west},
      grid style = {very thin, dashed, gray!40},
      grid = major,
      xmin=1980, xmax=2021,
      ymin=0, ymax=8,
      xlabel = {year},
      ylabel = {homicides per 100,000 people}]

      \addplot+ [name path=lower, fill=none, mark=none, draw=none, forget plot] table [
        x=year, y=lower, col sep=comma] {IHME-GBD_2021_DATA-e74d460c-1.csv};
      \addplot+ [name path=upper, fill=none, mark=none, draw=none, forget plot] table [
        x=year, y=upper, col sep=comma] {IHME-GBD_2021_DATA-e74d460c-1.csv};
      \addplot+ [fill opacity=0.1, forget plot] fill between[of=lower and upper];

      \addplot+ [mark size=0.3] 
        table[x=year,y=val,col sep=comma]
             {IHME-GBD_2021_DATA-e74d460c-1.csv}; 
      \addlegendentry{Wrld}  

      \addplot+ [mark size=0.3] 
        table[x=Year,y=Rate,col sep=comma]
             {homicide-rates-across-western-europe-1.csv}; 
      \addlegendentry{CoSa}

      \addplot+ [mark size=0.3] 
        table[x=Year,y=Rate,col sep=comma]
             {homicide-rates-across-western-europe-3.csv};  
      \addlegendentry{EnWa}

      \addplot+ [mark size=0.3] 
        table[x=Year,y=Rate,col sep=comma]
             {homicide-rates-across-western-europe-4.csv};
      \addlegendentry{BEL}

      \addplot+ [mark size=0.3] 
        table[x=Year,y=Rate,col sep=comma]
             {homicide-rates-across-western-europe-5.csv};
      \addlegendentry{FRA}

      \addplot+ [mark size=0.3] 
        table[x=Year,y=Rate,col sep=comma]
             {homicide-rates-across-western-europe-6.csv};
      \addlegendentry{NLD}

      \addplot+ [mark size=0.3] 
        table[x=Year,y=Rate,col sep=comma]
             {homicide-rates-across-western-europe-7.csv};
      \addlegendentry{ESP}

      \addplot+ [mark size=0.3] 
        table[x=Year,y=Rate,col sep=comma]
             {homicide-rates-across-western-europe-8.csv};
      \addlegendentry{DEU}

      \addplot+ [mark size=0.3] 
        table[x=Year,y=Rate,col sep=comma]
             {homicide-rates-across-western-europe-9.csv};
      \addlegendentry{CHE}

      \addplot+ [mark size=0.3] 
        table[x=Year,y=Rate,col sep=comma]
             {homicide-rates-across-western-europe-10.csv};
      \addlegendentry{ITA}

      \addplot+ [mark size=0.3] 
        table[x=Year,y=Rate,col sep=comma]
             {homicide-rates-across-western-europe-11.csv};
      \addlegendentry{IRL}

    \end{axis}
  \end{tikzpicture}

  \vspace{-0.55em}
  \caption[Homicide rates, 1980 to 2021.]
          {\hspace{-0.40em}\hyperlink{raceFigRef}{\(^{\uparrow}\)}\hspace{0.41em}Homicide
           rates, 1980 to 2021.
           A zoom-in of figure \ref{fig:homicide}.
           Non-``Wrld'' data are from \cite{owidHomicideLongTerm};
           for original sources, see
           \cite{owidHomicideLongTerm}.
           Data for ``Wrld'' (world) are from \cite{ihme};
           see also \cite{owidHomicideRate}.
           See \cite{owidHomicideDataSources} too.
           `CoSa': Corsica and Sardinia,
           `EnWa': England and Wales.
          }
  \label{fig:homicideRecent}
  \vspace{2mm}
\end{figure}

\begin{figure}
  \centering

  \begin{tikzpicture}

    \begin{axis}[
      legend style={font=\footnotesize},
      legend style={fill=none},  
      legend cell align=left,
      width = 0.67\linewidth,
      height = 0.80\linewidth,
      axis y line = left,
      axis x line = bottom,
      axis line style = {very thick},
      xticklabel style={/pgf/number format/1000 sep={}},
      yticklabel={$\pgfmathprintnumber{\tick}\%$},
      legend style={at={(axis cs:2023,20.0)},anchor=south west},
      grid style = {thin, densely dotted, black!50},
      xmin=1984, xmax=2022,
      ymin=0, ymax=80,
      xlabel = {year},
      ylabel = {share saying ``most can be trusted''}]

      \addplot+ [mark size=0.3] 
        coordinates {(1984,52.09644)(1993,59.59885)(1998,56.59068)(2004,63.71845)(2010,63.94534)(2014,60.10379)(2022,62.76585)};
      \addlegendentry{\texttt{SWE 10.7\,pp}}

      \addplot+ [mark size=0.3] 
        coordinates {(1984,40.5972)(1993,41.73789)(2004,39.2562)(2010,48.51524)(2022,62.27)};
      \addlegendentry{\texttt{ISL 21.7\,pp}}

      \addplot+ [mark size=0.3] 
        coordinates {(1984,37.9408905)(2010,40.105393)(2022,43.479468)};
      \addlegendentry{\texttt{WLD \ 5.5\,pp}}

      \addplot+ [mark size=0.3] 
        coordinates {(1984,42.48064)(1993,42.13953)(1998,30.44983)(2004,28.54928)(2010,35.34413)(2022,43.31179)};
      \addlegendentry{\texttt{GBR \ 0.8\,pp}}

      \addplot+ [mark size=0.3] 
        coordinates {(1984,32.23605)(1993,32.063)(1998,28.66726)(2004,34.50539)(2010,27.64687)(2014,18.97492)(2022,40.96439)};
      \addlegendentry{\texttt{ESP \ 8.7\,pp}}
     
      \addplot+ [mark size=0.3] 
        coordinates {(1984,43.06532)(1993,59.52381)(1998,47.923)(2004,56.75382)(2010,59.9127)(2022,38.80566)};
      \addlegendentry{\texttt{NIR -4.3\,pp}}

      \addplot+ [mark size=0.3] 
        coordinates {(1984,39.2496)(1993,49.9909)(1998,34.95129)(2004,35.51853)(2010,39.13124)(2014,34.81074)(2022,37.02172)};
      \addlegendentry{\texttt{USA -2.2\,pp}}

      \addplot+ [mark size=0.3] 
        coordinates {(1984,31.91018)(23.82382)(1998,22.46154)(2004,21.40927)(2010,23.9954)(2022,27.15409)};
      \addlegendentry{\texttt{HUN -4.8\,pp}}

    \end{axis}
  \end{tikzpicture}

  \vspace{-0.55em}
  \caption[Share saying that most people can be trusted, 1984 to 2022.]
          {\hspace{-0.40em}\hyperlink{trustFigRef}{\(^{\uparrow}\)}\hspace{0.41em}Share
                 saying that most people can be trusted, 1984 to 2022,
           for all 20 countries with entries for both 1984 and 2022.
           Wld means average of these 20 countries. 
           Legend also shows percentage point difference between 2022 and 1984.
           Data from IVS (EVS and WVS \cite{WVS}), via \cite{owidTrust}. 
           See \cite{owidTrust}.
           See figure \ref{fig:trust} for rest of countries.
          }
          
  \label{fig:trustB}

  \vspace{1.00em}
\end{figure}


\begin{figure}[!htb]
  \centering

  \begin{tikzpicture}

    \begin{axis}[
      legend style={font=\footnotesize},
      legend style={fill=none},  
      legend cell align=left,
      width = 0.80\linewidth,
      height = 1.20\linewidth,
      axis y line = left,
      axis x line = bottom,
      axis line style = {very thick},
      xticklabel style={/pgf/number format/1000 sep={}},
      legend style={at={(axis cs:2024,1.00)},anchor=south west},
      grid style = {very thin, dashed, gray!40},
      grid = major,
      xmin=1950, xmax=2023,
      ymin=0, ymax=7,
      xlabel = {year},
      ylabel = {births per woman}]

      \addplot+ [mark size=0.6] 
        table[x=Year,y=Fertility_rate,col sep=comma]
             {children-born-per-woman-nga.csv};
      \addlegendentry{Nigeria}

      \addplot+ [mark size=0.6] 
        table[x=Year,y=Fertility_rate,col sep=comma]
             {children-born-per-woman-000-afr.csv}; 
      \addlegendentry{Africa}

      \addplot+ [mark size=0.6] 
        table[x=Year,y=Fertility_rate,col sep=comma]
             {children-born-per-woman-111-wrld.csv};
      \addlegendentry{World}

      \addplot+ [mark size=0.3] 
        table[x=Year,y=Fertility_rate,col sep=comma]
             {children-born-per-woman-101-oceania.csv};
      \addlegendentry{Oceania}

      \addplot+ [mark size=0.3] 
        table[x=Year,y=Fertility_rate,col sep=comma]
             {children-born-per-woman-000-asia.csv};
      \addlegendentry{Asia}

      \addplot+ [mark size=0.3] 
        table[x=Year,y=Fertility_rate,col sep=comma]
             {children-born-per-woman-110-latam.csv};
      \addlegendentry{Lat Amer}

      \addplot+ [mark size=0.2] 
        table[x=Year,y=Fertility_rate,col sep=comma]
             {children-born-per-woman-001-na.csv};
      \addlegendentry{N Amer}

      \addplot+ [green,mark size=0.3] 
        table[x=Year,y=Fertility_rate,col sep=comma]
             {children-born-per-woman-100-europe.csv};
      \addlegendentry{Europe}

      \addplot+ [mark size=0.3] 
        table[x=Year,y=Fertility_rate,col sep=comma]
             {children-born-per-woman-011-esp.csv};
      \addlegendentry{Spain}

    \end{axis}
  \end{tikzpicture}

  \vspace{-0.55em}
  \caption[Human fertility rates, 1950 to 2023.]
          {\hspace{-0.78em}\hyperlink{humanFertBFigRef}{\(^{\uparrow}\)}\hspace{0.79em}Human
           fertility rates, 1950 to 2023.
           Data from \cite{UNFert}, via \cite{owidfertilityrate}.
           `Lat Amer': Latin America and the Caribbean.
           See figures \ref{fig:humanFertility} and \ref{fig:humanFertilityC}
           for other countries and constellations.}
  \label{fig:humanFertilityB}
  \vspace{-0.10em}
\end{figure}

\begin{figure}[!htb]
  \centering

  \begin{tikzpicture}

    \begin{axis}[
      legend style={font=\footnotesize},
      legend cell align=left,
      width = 1.05\linewidth,
      height = 1.55\linewidth,
      axis y line = left,
      axis x line = bottom,
      axis line style = {very thick},
      xticklabel style={/pgf/number format/1000 sep={}},
      legend style={at={(axis cs:2001,4.1)},anchor=south west},
      grid style = {very thin, dashed, gray!40},
      grid = major,
      xmin=1950, xmax=2023,
      ymin=0, ymax=8,
      xlabel = {year},
      ylabel = {births per woman}]

      \addplot+ [mark size=0.6] 
        table[x=Year,y=Fertility_rate,col sep=comma]
             {children-born-per-woman-egy.csv};
      \addlegendentry{Egypt}

      \addplot+ [mark size=0.5] 
        table[x=Year,y=Fertility_rate,col sep=comma]
             {children-born-per-woman-111-wrld.csv};
      \addlegendentry{World}

      \addplot+ [densely dashed,mark size=0.6] 
        table[x=Year,y=Fertility_rate,col sep=comma]
             {children-born-per-woman-010-india.csv};
      \addlegendentry{India}

      \addplot+ [mark size=0.3] 
        table[x=Year,y=Fertility_rate,col sep=comma]
             {children-born-per-woman-iran.csv};
      \addlegendentry{Iran}

      \addplot+ [mark size=0.3] 
        table[x=Year,y=Fertility_rate,col sep=comma]
             {children-born-per-woman-tur.csv};
      \addlegendentry{Turkey}

      \addplot+ [mark size=0.3] 
        table[x=Year,y=Fertility_rate,col sep=comma]
             {children-born-per-woman-bra.csv};
      \addlegendentry{Brazil}


      \addplot+ [green,densely dotted,mark size=0.4] 
        table[x=Year,y=Fertility_rate,col sep=comma]
             {children-born-per-woman-101-rus.csv};
      \addlegendentry{Russia}

      \addplot+ [mark size=0.3] 
        table[x=Year,y=Fertility_rate,col sep=comma]
             {children-born-per-woman-ury.csv};
      \addlegendentry{Uruguay}

      \addplot+ [mark size=0.3] 
        table[x=Year,y=Fertility_rate,col sep=comma]
             {children-born-per-woman-jap.csv};
      \addlegendentry{Japan}

      \addplot+ [densely dotted,mark size=0.3] 
        table[x=Year,y=Fertility_rate,col sep=comma]
             {children-born-per-woman-uae.csv};
      \addlegendentry{UAE}

      \addplot+ [black,dotted,mark size=0.4] 
        table[x=Year,y=Fertility_rate,col sep=comma]
             {children-born-per-woman-000-china.csv};
      \addlegendentry{China}

    \end{axis}
  \end{tikzpicture}

  \vspace{-0.60em}
  \caption[Human fertility rates, 1950 to 2023.]
          {\hspace{-0.44em}\hyperlink{humanFertBFigRef}{\(^{\uparrow}\)}\hspace{0.45em}Human
           fertility rates, 1950 to 2023.
           Data from \cite{UNFert}, via \cite{owidfertilityrate}.
           See figures \ref{fig:humanFertility} and \ref{fig:humanFertilityB}
           for other countries and constellations.}
  \label{fig:humanFertilityC}
  \vspace{0.60em}
\end{figure}

\begin{figure}
  \centering

  \begin{tikzpicture}

    \begin{axis}[
      ybar,
      legend style={at={(0.05,1.03)},
        anchor=south west},
      symbolic x coords={``religious'',``non-religious'',``atheist''},
      xtick=data,
      legend columns=3,
      legend style={font=\footnotesize},
      width = 1.00\linewidth,
      height = 0.81\linewidth,
      bar width= 5pt,
      ybar=2pt,
      yticklabel={$\pgfmathprintnumber{\tick}\%$},
      grid style = {very thin, dashed, gray!40},
      minor y tick num=3,
      ymajorgrids=true,
      yminorgrids=true,
      xmajorgrids=false,
      xminorgrids=false,
      enlarge x limits={abs=1.2cm},
      extra y ticks={10,30,50,70,90},
      ymin=0, ymax=70,
      ylabel = {share mentioning quality as important in child}
    ]


    \addplot +[pattern=north west lines,
               error bars/.cd, y dir=both, y explicit] coordinates {
      (``religious'',42.2877) +- (``religious'',0.3153)
      (``non-religious'',53.6763) +- (``non-religious'',0.4729)
      (``atheist'',62.2497) +- (``atheist'',0.8306)
    };
    \addlegendentry{independence}

    \addplot +[error bars/.cd, y dir=both, y explicit] coordinates {
      (``religious'',18.0209) +- (``religious'',0.2458)
      (``non-religious'',26.9570) +- (``non-religious'',0.4219)
      (``atheist'',34.3807) +- (``atheist'',0.8150)
    };
    \addlegendentry{imagination}

    \addplot +[pattern=dots,
               error bars/.cd, y dir=both, y explicit] coordinates {
      (``religious'',30.2572)  +- (``religious'',0.2936)
      (``non-religious'',23.0728) +- (``non-religious'',0.4007)
      (``atheist'',13.2613) +- (``atheist'',0.5818)
    };
    \addlegendentry{obedience}

    \addplot +[error bars/.cd, y dir=both, y explicit] coordinates {
      (``religious'',42.6467)  +- (``religious'',0.3151)
      (``non-religious'',56.1825) +- (``non-religious'',0.4697)
      (``atheist'',69.2854) +- (``atheist'',0.7900)
    };
    \addlegendentry{binary composite emancipative measure\hspace*{-11.5em}}

    \end{axis}
  \end{tikzpicture}

  \vspace{-0.55em}
  \caption[Parts of a composite emancipative measure.]
          {\hspace{-0.73em}\hyperlink{emancFigRef}{\(^{\uparrow}\)}\hspace{0.74em}Parts
           of a composite emancipative measure, for each
           of the three disjoint groups. 
           Data from IVS (EVS and WVS),\cite{ivs2022} with non-answers discarded.
           All 92 countries included in the 2017-2022 survey are aggregated,
           on a per-respondent basis.
           95\% confidence intervals. The number of qualities a respondent can list
           is limited. Mentioning obedience should lower the emancipative
           measure. See \cite{emancRef,welzel13}.$^{\ref{fn:ivsExt}}$
           Also shown is a per-respondent binary composite average of the three 
           qualities (with the obedience value inverted).\footnotemark{}}
          
  \label{fig:emanc}

  \vspace{-0.30em}
\end{figure}

\if 01

A029 - Important child qualities: independence
./ivsExtract -f EVS_WVS_Joint_Csv_v5_0.csv -F 6 -p 1:166 -a 1:4 -p 2:52 -a 2:3 -c 0 -v9 -m 2
poi column:  52 "A029"
Results with non-answers discarded:
 x y: count  share     count / total  +- MoE
 1 1: 54410  0.577123  54410 / 94278  0.003153
 1 2: 39868  0.422877  39868 / 94278  0.003153
 2 1: 19783  0.463237  19783 / 42706  0.004729
 2 2: 22923  0.536763  22923 / 42706  0.004729
 3 1:  4940  0.377503   4940 / 13086  0.008306
 3 2:  8146  0.622497   8146 / 13086  0.008306

A034 - Important child qualities: imagination
/ivsExtract -f EVS_WVS_Joint_Csv_v5_0.csv -F 6 -p 1:166 -a 1:4 -p 2:55 -a 2:3 -c 0 -v9 -m 2
poi column:  55 "A034"
Results with non-answers discarded:
 x y: count  share     count / total  +- MoE
 1 1: 77021  0.819791  77021 / 93952  0.002458
 1 2: 16931  0.180209  16931 / 93952  0.002458
 2 1: 31044  0.730430  31044 / 42501  0.004219
 2 2: 11457  0.269570  11457 / 42501  0.004219
 3 1:  8562  0.656193   8562 / 13048  0.008150
 3 2:  4486  0.343807   4486 / 13048  0.008150

A042 - Important child qualities: obedience
./ivsExtract -f EVS_WVS_Joint_Csv_v5_0.csv -F 6 -p 1:166 -a 1:4 -p 2:61 -a 2:3 -c 0  -v5 -m 2
poi column:  61 "A042"
poi column: 166 "F034"
Rows: 156659
Rows with investigated country: 156658
Results with non-answers discarded:
 x y: count  share     count / total  +- MoE
 1 1: 65575  0.697428  65575 / 94024  0.002936
 1 2: 28449  0.302572  28449 / 94024  0.002936
 2 1: 32671  0.769272  32671 / 42470  0.004007
 2 2:  9799  0.230728   9799 / 42470  0.004007
 3 1: 11322  0.867387  11322 / 13053  0.005818
 3 2:  1731  0.132613   1731 / 13053  0.005818

composite emancipative measure:
Results with non-answers discarded:
 x y: count  share     count / total  +- MoE
 1 1: 54274  0.573533  54274 / 94631  0.003151
 1 2: 40357  0.426467  40357 / 94631  0.003151
 2 1: 18785  0.438175  18785 / 42871  0.004697
 2 2: 24086  0.561825  24086 / 42871  0.004697
 3 1:  4023  0.307146   4023 / 13098  0.007900
 3 2:  9075  0.692854   9075 / 13098  0.007900

generated using:
./ivsExtract -f EVS_WVS_Joint_Csv_v5_0.csv -F 6 -B 3 -D 2 -p 1:166 -a 1:4 -p 3:52 -p 4:55 -p 5:61 -a 3:3 -a 4:3 -a 5:3 -c 0 -v9 -m 3 -m 4 -m 5 -O 1 -a 2:3

\fi

%
%
\footnotetext{For
\label{fn:ivsExtEmanc}
example, specifically, the data for the binary composite emancipative measure were generated using the 
command:\\ 
\textsf{./ivsExtract -f EVS\_WVS\_Joint\_Csv\_v5\_0.csv -B 3 -D 2 -p 1:166 -a 1:4 -p 3:52 -p 4:55 -p 5:61 -c 1 -m 3 -m 4 -m 5 -O 1 -a 2:3}}

\end{document}